\documentclass[aps,prd,twocolumn,superscriptaddress,nofootinbib,amsmath,amssymb,floatfix]{revtex4-2}

\usepackage{graphicx}
\usepackage{bm}
\usepackage{dcolumn}
\usepackage{xcolor}
\usepackage{mathbbol} 
\usepackage{tikz}
\usetikzlibrary{decorations.pathmorphing,shapes.geometric,arrows.meta,calc}
\usepackage[colorlinks=true,linkcolor=blue!50!black,citecolor=blue!50!black,urlcolor=blue!50!black]{hyperref}

\newif\ifmarkchanges
\markchangestrue
\newif\ifmarkprev
\markprevfalse

\newcommand{\addedprev}[1]{\ifmarkprev{\leavevmode\color{red}\hypersetup{linkcolor=red,citecolor=red}#1}\else\leavevmode#1\fi}

\newcommand{\Hb}{\bar H}
\newcommand{\Db}{\bar D}
\newcommand{\Dbs}{\bar D^{*}}
\newcommand{\Sc}{\Sigma_c}
\newcommand{\Scs}{\Sigma_c^{*}}
\newcommand{\Lc}{\Lambda_c}
\newcommand{\Jpsi}{J/\psi}
\newcommand{\etac}{\eta_c}
\newcommand{\Pc}{P_c}
\newcommand{\tr}[1]{\langle #1\rangle}
\newcommand{\vq}{\bm{q}}

\newcommand{\vP}{\bm{P}}
\newcommand{\vs}{\bm{\sigma}}
\newcommand{\vB}{\bm{B}}

\newcommand{\vE}{\bm{E}}
\newcommand{\vJ}{\bm{\mathcal J}}
\newcommand{\dd}{\mathrm d}
\newcommand{\rs}{\sqrt{s}}
\newcommand{\rsp}{\sqrt{s'}}
\newcommand{\one}{\mathbb{1}}
\newcommand{\muN}{\mu_N}
\newcommand{\GeV}{\,\mathrm{GeV}}
\newcommand{\MeV}{\,\mathrm{MeV}}
\newcommand{\fm}{\,\mathrm{fm}}

\newcommand{\Rl}{\operatorname{Re}}
\renewcommand{\Im}{\operatorname{Im}}

\newcommand{\Gt}{\Gamma}
\newcommand{\GE}{G_{E}}
\newcommand{\GM}{G_{M}}
\newcommand{\FB}{F_{B}}
\newcommand{\Fcb}{F_{\bar c}}
\newcommand{\FC}{F_{C}}
\newcommand{\EM}{\mathbb E^{\rm M}}
\newcommand{\EB}{\mathbb E^{\rm B}}
\newcommand{\MM}{\mathbb M^{\rm M}}
\newcommand{\MB}{\mathbb M^{\rm B}}
\newcommand{\Ean}{\mathbb E^{\rm an}}
\newcommand{\Ecal}{\mathcal E}
\newcommand{\mured}{\mu_{\rm red}}
\newcommand{\mst}{m_{\rm st}}
\newcommand{\msp}{m_{\rm sp}}
\newcommand{\Xis}{\Xi}
\newcommand{\MSR}{\langle r^{2}\rangle}
\newcommand{\MSD}{\langle x^{2}\rangle}
\newcommand{\zR}{z_R}
\newcommand{\Fan}{F^{\rm an}}
\newcommand{\Fcon}{F^{\rm con}}

\begin{document}

\title{Electromagnetic form factors of the hidden-charm pentaquarks in a coupled-channel unitary approach}

\author{Nongnapat Ponkhuha}
\email{Nongnapat.po@kkumail.com}
\author{Ratirat Suntharawirat}
\email{ratirat\_su@kkumail.com}
\author{Daris Samart}
\thanks{Corresponding author}
\email{darisa@kku.ac.th}
\affiliation{Khon Kaen Particle Physics and Cosmology Theory Group (KKPaCT),
Department of Physics, Faculty of Science, Khon Kaen University,
123 Mitraphap Rd., Khon Kaen 40002, Thailand}

\date{\today}

\begin{abstract}
We study the electromagnetic form factors of the hidden-charm pentaquarks $P_c(4312)$, $P_c(4440)$ and
$P_c(4457)$ and of their four heavy-quark spin partners, which are treated as $\Db^{(*)}\Sc^{(*)}$ hadronic
molecules. The states are generated as poles of a coupled-channel unitary amplitude with a vector-meson exchange
kernel that respects heavy-quark spin symmetry. The form factors are obtained from the residues at the double
pole of the photon-coupled scattering amplitude. The electromagnetic current consists of one-body currents and a
photon--four-hadron vertex, which we construct for the heavy-quark spin symmetric channel space. The
gauged-kernel part of this vertex ensures the charge normalization through the compositeness sum rule, while its
magnetic and electric contact parts contain couplings that are not determined by the model. For the spin
assignment $J^P=\tfrac12^-$, $\tfrac12^-$, $\tfrac32^-$ and pointlike constituents, the complex mean squared
distances between the constituents of the three observed states are $0.85-0.47i$, $0.35-0.18i$ and
$1.76-0.57i\fm^2$. Within $2\%$, these values are given by the zero-range result $1/(4\mured B_c)$, where
$\mured$ is the reduced mass and $B_c$ the complex binding energy, multiplied by the effective-range factor of
the model amplitude. The real parts of the magnetic moments are $1.7$, $-0.9$ and $1.3\,\muN$, which differ from
the additive moments of the constituents by at most $11\%$. Varying the constituent moments and the magnetic
contact couplings within bounds from dimensional analysis, we obtain the ranges $0.8$ to $2.0$, $-1.3$ to $-0.2$
and $0.2$ to $1.8\,\muN$. The charge radii depend in addition on the structure of the constituents and on
electric contact currents. We also show that the photon loops of open channels have a triangle singularity at
spacelike momentum transfer. The analytic continuation of the residue in $Q^2$ is specified, and the difference
between its two adjacent branches is given in closed form.
\end{abstract}

\maketitle

\section{Introduction}
\label{sec:intro}

The LHCb Collaboration observed hidden-charm pentaquark candidates in the $\Jpsi p$ invariant mass distribution of
$\Lambda_b^0\to\Jpsi pK^-$ decays~\cite{Aaij:2015tga}. With an enlarged data sample, three narrow states,
$P_c(4312)^+$, $P_c(4440)^+$ and $P_c(4457)^+$, were resolved~\cite{Aaij:2019vzc}, and a candidate with
strangeness was later observed in the $\Jpsi\Lambda$ system~\cite{LHCb:2023}. The three nonstrange states lie
$9$, $22$ and $5\MeV$ below the $\Db\Sc$ and $\Dbs\Sc$ thresholds. This proximity suggests that they are
hadronic molecules composed of an anticharmed meson and a charmed
baryon~\cite{Wu:2010jy,Wu:2010vk,Guo:2017jvc,Chen:2016qju,Liu:2019tjn,Xiao:2019aya}, although kinematical
explanations of part of the signals have also been
proposed~\cite{Guo:2015umn,Liu:2015fea,Mikhasenko:2015vca,Guo:2016bkl,Bayar:2016ftu,Nakamura:2021qvy}. Heavy-quark
spin symmetry (HQSS) relates the $\Db\Sc$, $\Db\Scs$, $\Dbs\Sc$ and $\Dbs\Scs$ systems and predicts seven
$S$-wave states with isospin $\tfrac12$~\cite{Liu:2019tjn,Xiao:2013yca,Xiao:2019aya}. The spin and parity of the
observed states have not yet been determined experimentally. Models with contact or vector-exchange
interactions assign $J^P=\tfrac12^-$ to $P_c(4440)$ and $\tfrac32^-$ to $P_c(4457)$~\cite{Xiao:2019aya}, whereas
analyses that include one-pion exchange and the observed $\Jpsi p$ spectrum prefer the opposite
ordering~\cite{Du:2020,Du:2021,Yamaguchi:2020}.

Electromagnetic properties provide direct information on the internal structure of hadrons. The magnetic
moments of the $P_c$ states have been calculated in the constituent quark model for molecular and compact
configurations~\cite{Wang:2016dzu,OrtizPacheco:2019,GuoLi:2024,LiHS:2024b}, in a molecular picture with coupled
channels and $D$-wave components~\cite{Li:2021ryu}, and in chiral perturbation theory~\cite{LiHS:2024a}. The
magnetic dipole and higher multipole moments have also been studied with light-cone QCD sum rules for molecular
and compact interpolating currents~\cite{Ozdem:2018qeh,Ozdem:2021,Ozdem:2024,Ozdem:2024iso,Ozdem:2026cmp}, and
the strange partners have been investigated in Refs.~\cite{WangFL:2022,Ozdem:2026}. The radiative transition
$P_c(4457)\to P_c(4312)\gamma$ has been studied in Ref.~\cite{Suntharawirat:2026}. To the best of our knowledge,
however, the momentum dependence of the electromagnetic form factors of the nonstrange $P_c$ states at their
pole positions has not been studied so far.

In this work, we calculate the electromagnetic form factors of the seven states generated in the
coupled-channel model with the HQSS vector-exchange kernel of Ref.~\cite{Xiao:2019aya}. We follow the method
developed by Sekihara, Hyodo and Jido for the $\Lambda(1405)$~\cite{Sekihara:2008qk,Sekihara:2010uz}, in which
the form factor of a resonance is defined by the residue at the double pole of the photon-coupled scattering
amplitude. The same definition is used in the on-shell formalism of Ref.~\cite{Briceno:2020vgp}. In this
approach, no ansatz for the wave function is required. Compared with the $\Lambda(1405)$, the present system
has three distinct features. First, the $P_c^+$ is charged, so that the photon vertex generated by gauging the
energy-dependent kernel contributes to the current. This vertex provides the energy-dependence term of the
compositeness sum rule~\cite{Hyodo:2011qc,Aceti:2012dd,Sekihara:2014kya} in the charge normalization. Second,
the coupled channels are related by HQSS, and the magnetic operators connect different channels. For this
channel space, we construct the photon--four-hadron vertex, which consists of the gauged kernel and of the
lowest magnetic and electric contact operators allowed by HQSS. Third, decay channels open slightly below the
poles. The photon loops of these channels have a triangle singularity at spacelike momentum transfer, which
requires a specification of the analytic continuation of the residue in $Q^2$.

This paper is organized as follows. In Sec.~\ref{sec:formalism}, we define the form factors and introduce the
coupled-channel model and the photon couplings. In Sec.~\ref{sec:poles}, we present the form factors and radii
and discuss their relation to the zero-range limit. The magnetic moments and the two spin orderings are
discussed in Sec.~\ref{sec:moments}. The sensitivity of the results to the observed masses and widths is examined
in Sec.~\ref{sec:measured}, and the open channels are treated in Sec.~\ref{sec:open}. Section~\ref{sec:summary}
is devoted to a summary. Radii are given in fm$^2$ with $\hbar c=197.327\MeV\fm$.

\section{Formalism}
\label{sec:formalism}

\subsection{Form factors of a resonance}
\label{subsec:def}

Throughout this paper, we use the metric $(+,-,-,-)$, natural units $\hbar=1=c$, $e=\sqrt{4\pi\alpha}$, $\alpha=1/137$ and
$\mathcal L_{\rm int}=-eJ^\mu_{\rm em}A_\mu$. The virtual photon is incoming in the process $\Pc(P)+\gamma^*(q)\to\Pc(P')$,
with $q=P'-P$ and $Q^2=-q^2$. Following Refs.~\cite{Sekihara:2008qk,Sekihara:2010uz}, we define the form factors
of a resonance with the pole position $\zR=M_R-i\Gamma_R/2$ by the matrix element of the current between the
resonance states,
\begin{equation}
\langle\zR(\vP')^{(-)}|J^\mu(x)|\zR(\vP)^{(+)}\rangle=e^{i(P'-P)\cdot x}\,F^\mu(Q^2),
\label{eq:FFdef}
\end{equation}
which is complex in general. In the Breit frame, where $\vP=-\vq/2$ and $\vP'=+\vq/2$, the charge and magnetic
dipole form factors are defined with nonrelativistic spin wave functions as
\begin{equation}
\begin{aligned}
\langle J^0\rangle_{M'M}&=\GE(Q^2)\,\delta_{M'M},\\
\langle\bm J\rangle_{M'M}&=\frac{i}{2M_p}\,\frac{\GM(Q^2)}{J}\,\big(\vJ\times\vq\big)_{M'M},
\end{aligned}
\label{eq:GEGM}
\end{equation}
where $\vJ$ are the spin-$J$ matrices, $M_p$ is the proton mass and the current is given in units of $e$. Both
form factors are dimensionless and normalized as $\GE(0)=1$ and $\GM(0)=\mu/\muN$ with $\muN=e/(2M_p)$. They
coincide with the covariant form factors up to corrections of relative order $Q^2/(4M_R^2)$, which amount to
$1.3\%$ at $Q^2=1\GeV^2$. The electric quadrupole and magnetic octupole form factors of the states with
$J\geq\tfrac32$ require $D$-wave components or the higher multipoles of the $\Dbs$ and $\Scs$, and are therefore
not included in the present $S$-wave model. Nonvanishing higher multipole moments of the spin-$\tfrac32$ states
have been obtained with light-cone sum rules~\cite{Ozdem:2024iso}.

In addition to the electromagnetic current, we consider three probe currents: the baryon-number current ($\FB$),
the current that counts the anticharm quarks ($\Fcb$) and the charm current $J^\mu_C=J^\mu_{N_c}-J^\mu_{N_{\bar
c}}$ ($\FC$). For pointlike constituents, $\FB$ traces the position of the baryon and $\Fcb$ that of the meson.
The mean squared radius associated with a probe is $\MSR=-6\,\dd F/\dd Q^2$ at $Q^2=0$, divided by $\GM(0)$ in
the case of the magnetic radius.

We denote by $T_{l'l}(\rs)$ the $S$-wave meson--baryon amplitude for $M_lB_l\to M_{l'}B_{l'}$. Near a pole on the
appropriate Riemann sheet, $T_{l'l}\simeq g_{l'}g_l/(\rs-\zR)$, where the residues define the effective couplings
$g_l$ of the state to the channels~\cite{Jido:2002yz}. In the photon-coupled amplitude
$T^\mu_{\gamma,l'l}(\rsp,\rs;Q^2)$ for $M_lB_l+\gamma^*\to M_{l'}B_{l'}$, the resonance appears as a double pole,
and the matrix element \eqref{eq:FFdef} is given by its
coefficient~\cite{Sekihara:2008qk,Sekihara:2010uz,Briceno:2020vgp},
\begin{equation}
F^\mu(Q^2)=\frac{(\rsp-\zR)\,T^\mu_{\gamma,l'l}(\rsp,\rs;Q^2)}{T_{l'l}(\rs)}\Big|_{\rs,\rsp\to\zR} .
\label{eq:residue}
\end{equation}
The nonresonant background and all terms with at most a single pole do not contribute to Eq.~\eqref{eq:residue}.
The amplitudes are normalized as $i\mathcal M_\gamma=-ie\,\epsilon_\mu T^\mu_\gamma$ and $T=-\mathcal M$. With our
sign convention for $\mathcal L_{\rm int}$, the amplitude $T_\gamma$ differs in overall sign from that of
Ref.~\cite{Sekihara:2010uz}, whereas the form factors are identical.

\subsection{Coupled-channel model}
\label{subsec:model}

We employ the isospin-$\tfrac12$ coupled channels of Ref.~\cite{Xiao:2019aya},
\begin{equation}
\begin{aligned}
\tfrac12^-:&\ \{\etac N,\,\Jpsi N,\,\Db\Lc,\,\Db\Sc,\,\Dbs\Lc,\,\Dbs\Sc,\,\Dbs\Scs\},\\
\tfrac32^-:&\ \{\Jpsi N,\,\Dbs\Lc,\,\Dbs\Sc,\,\Db\Scs,\,\Dbs\Scs\},\\
\tfrac52^-:&\ \{\Dbs\Scs\},
\end{aligned}
\label{eq:channels}
\end{equation}
which are labeled by $l$, with the meson mass $m_l$ and the baryon mass $M_l$. The isospin states are
$|\Db^{(*)}\Sc^{(*)}\rangle=\sqrt{2/3}\,|D^{(*)-}\Sc^{(*)++}\rangle-\sqrt{1/3}\,|\Db^{(*)0}\Sc^{(*)+}\rangle$ and
$|\Db^{(*)}\Lc\rangle=|\Db^{(*)0}\Lc^+\rangle$. We use the isospin-averaged masses of
Ref.~\cite{ParticleDataGroup:2026} (in MeV: $\etac$ 2984.1, $\Jpsi$ 3096.9, $\Db$ 1867.25, $\Dbs$ 2008.56, $N$
938.92, $\Lc$ 2286.46, $\Sc$ 2453.46, $\Scs$ 2518.1).

In the heavy-quark limit, the interaction depends only on the light degrees of freedom. The most general
$S$-wave kernel $V_{l'l}(J)$ consistent with HQSS is given in Eqs.~(3)--(5) of Ref.~\cite{Xiao:2019aya} in terms
of seven couplings, $\mu_1$, $\mu_2$, $\mu_3$, $\mu_{12}$, $\mu_{13}$ and $\mu_{23}$ for light-quark spin
$\tfrac12$ and $\lambda_2$ for light-quark spin $\tfrac32$~\cite{GarciaRecio:2013gaa,Xiao:2013yca}. These
couplings are determined by the extended local hidden gauge approach~\cite{Wu:2010jy,Wu:2010vk,Xiao:2013yca}, in
which the interaction arises from the exchange of $\rho$ and $\omega$ mesons between the light quarks and from
$D^*$ exchange in the transitions to the hidden-charm channels. Following Eq.~(6) of Ref.~\cite{Xiao:2019aya},
we use
\begin{equation}
\begin{aligned}
&\mu_1=\mu_{23}=0,\quad \mu_2=-\mu_3=\frac{k^0_{l'}+k^0_l}{4f^2},\\
&\lambda_2=r\,\mu_3,\qquad \mu_{13}=-\mu_{12},\\
&\mu_{12}=-\sqrt6\,\kappa\,\frac{m_\rho^2}{p_{D^*,l'l}^2-m_{D^*}^2}\,\frac{k^0_{l'}+k^0_l}{4f^2},
\end{aligned}
\label{eq:HGvalues}
\end{equation}
where two additional parameters $r$ and $\kappa$ have been introduced, $f=93\MeV$, $m_\rho=775.5\MeV$,
$k^0_l=(s+m_l^2-M_l^2)/(2\rs)$ is the meson energy and $p^2_{D^*,l'l}=m_{l'}^2+m_l^2-2k^0_{l'}k^0_l$~\cite{Wu:2010vk}.
The hidden gauge approach corresponds to $r=\kappa=1$, whereas HQSS alone does not fix these parameters. Since
the magnetic operators of Sec.~\ref{subsec:EMint} connect different channels, the relative phases of the channel
states affect the form factors. We therefore evaluate the kernel and all photon vertices in a single phase
convention, defined by the multiplet fields $\Hb=-\Db+\vs\cdot\bar{\bm D}{}^{*}$,
$S^i=\tfrac1{\sqrt3}\sigma^i\Sc+\Scs{}^i$ and $\Psi=-\etac+\vs\cdot\bm\psi$~\cite{Sakai:2019qph}. This convention
corresponds to the Condon--Shortley coupling scheme, in which the $\bar c$ is coupled first in the mesons, the
$c$ first in the baryons and the meson first in the two-hadron state. In this convention, we reproduce the
kernel matrices of Ref.~\cite{Xiao:2019aya}, including the relative sign $\mu_{13}=-\mu_{12}$ that enters the
interference terms of $\GM$.

Using the on-shell factorization of the kernel, the Bethe--Salpeter equation reduces to the algebraic
form~\cite{Oset:1997it,Oller:2000fj}
\begin{equation}
T(\rs)=\big[\one-V(\rs)G(\rs)\big]^{-1}V(\rs),\quad G={\rm diag}(G_l),
\label{eq:BSE}
\end{equation}
where $G_l$ is the meson--baryon loop function in dimensional regularization~\cite{Oller:2000fj},
\begin{equation}
\begin{split}
G_l(\rs)&=i\!\int\!\frac{\dd^4q}{(2\pi)^4}\,\frac{2M_l}{[(P-q)^2-M_l^2]\,[q^2-m_l^2]}\\
&=\frac{2M_l}{16\pi^2}\Big[a_l+2\\
&\quad+\!\int_0^1\!\!\dd x\ln\frac{xM_l^2+(1-x)m_l^2-x(1-x)s}{\mu_{\rm reg}^2}\Big],
\end{split}
\label{eq:Gloop}
\end{equation}
Here, $a_l$ is the subtraction constant at the regularization scale $\mu_{\rm reg}=1\GeV$, and
$s\to s+i\varepsilon$. The factor $2M_l$ of the baryon propagator corresponds to the normalization
$\bar uu=1$~\cite{Oset:1997it}. In this convention, $V$ and $T$ have dimension $E^{-1}$, $G$ has dimension $E$,
and the couplings $g_l$ are dimensionless. A negative $V$ corresponds to attraction. The loop function on the
second Riemann sheet of channel $l$ is given by
\begin{equation}
G^{\rm II}_l(\rs)=G^{\rm I}_l(\rs)+i\,\frac{M_lq_l}{2\pi\rs},\qquad \Im q_l>0 ,
\label{eq:GII}
\end{equation}
where $q_l$ is the center-of-mass momentum on the first sheet. We refer to a channel as open if its threshold
lies below $M_R$, and we search for the poles on the Riemann sheet on which all open channels are taken on
their second sheets.

In this work we consider two parameter sets as follows. \emph{Set I} corresponds to the model of Ref.~\cite{Xiao:2019aya}, with
$r=\kappa=1$ and a common subtraction constant $a_l=-2.09$. In this set, the two $\Dbs\Sc$ states lie within
$0.5\MeV$ of each other and deviate by $5$ and $12\MeV$ from the observed masses. In \emph{set II}, the
parameters are fitted to the observed masses and widths reported by the LHCb Collaboration~\cite{Aaij:2019vzc}. We identify these values with the pole positions. A fit of $a$, $r$ and $\kappa$ reproduces the widths, but the $\Db\Sc$ state
then lies $5.7\MeV$ below the observed mass of $P_c(4312)$, whose lower uncertainty is $0.9\MeV$. We therefore
allow the subtraction constant of the loops with a $\Db$ meson to differ from that of the loops with a $\Dbs$
meson, $a_{\Db}=a+\delta$, which corresponds to an effect of order $\Lambda_{\rm QCD}/m_c$. The three masses
determine $a$, $\delta$ and $r$, and $\kappa$ is fixed by a least-squares fit to the three widths, with the
statistical and systematic uncertainties added in quadrature. We obtain
\begin{equation}
\begin{aligned}
a&=-2.227,&\quad \delta&=+0.052,\\
r&=0.554,&\quad \kappa&=0.72^{+0.10}_{-0.14},
\end{aligned}
\label{eq:setII}
\end{equation}
with $\chi^2_{\min}=1.6$. The uncertainty of $\kappa$ is given by the interval $\chi^2\leq\chi^2_{\min}+1$, in
which the other parameters are redetermined for each value of $\kappa$. The value $r<1$ reduces the attraction
for light-quark spin $\tfrac32$, so that the $\tfrac12^-$ $\Dbs\Sc$ state lies below the $\tfrac32^-$ state.

\begin{table*}[t]
\caption{Pole positions of the seven states in the two parameter sets (masses and widths in MeV), compositeness
$X_{\rm main}$ of the dominant channel and energy-dependence term $Z$ of Eq.~\eqref{eq:sumrule} in set II, and
observed masses and widths~\cite{Aaij:2019vzc}. The states are labeled by their dominant channel. In set I,
$X_{\rm main}=0.99$--$1.01$ and $\Rl Z=0.006$--$0.013$.}
\label{tab:poles}
\begin{ruledtabular}
\begin{tabular}{lcccccccc}
 & \multicolumn{2}{c}{set I} & \multicolumn{4}{c}{set II} & \multicolumn{2}{c}{experiment}\\
state & $M_R$ & $\Gamma_R$ & $M_R$ & $\Gamma_R$ & $X_{\rm main}$ & $Z$ & $M$ & $\Gamma$\\
\hline
$\Db\Sc(\tfrac12^-)$ & $4306.6$ & $14.1$ & $4311.9$ & $10.1$ & $0.987-0.005i$ & $0.012-0.014i$ & $4311.9\pm0.7^{+6.8}_{-0.6}$ & $9.8\pm2.7^{+3.7}_{-4.5}$ \\
$\Db\Scs(\tfrac32^-)$ & $4374.4$ & $13.7$ & $4378.3$ & $10.8$ & $0.986-0.011i$ & $0.012-0.017i$ &  &  \\
$\Dbs\Sc(\tfrac12^-)$ & $4452.8$ & $23.5$ & $4440.3$ & $23.7$ & $0.977+0.009i$ & $0.014-0.030i$ & $4440.3\pm1.3^{+4.1}_{-4.7}$ & $20.6\pm4.9^{+8.7}_{-10.1}$ \\
$\Dbs\Sc(\tfrac32^-)$ & $4452.3$ & $3.0$ & $4457.3$ & $3.0$ & $0.986-0.009i$ & $0.010-0.002i$ & $4457.3\pm0.6^{+4.1}_{-1.7}$ & $6.4\pm2.0^{+5.7}_{-1.9}$ \\
$\Dbs\Scs(\tfrac12^-)$ & $4520.4$ & $22.3$ & $4505.4$ & $22.5$ & $0.987+0.018i$ & $0.013-0.026i$ &  &  \\
$\Dbs\Scs(\tfrac32^-)$ & $4519.0$ & $13.7$ & $4516.6$ & $12.7$ & $0.977+0.007i$ & $0.014-0.011i$ &  &  \\
$\Dbs\Scs(\tfrac52^-)$ & $4519.2$ & $0$ & $4524.6$ & $0$ & $0.994$ & $0.006$ &  &  \\
\end{tabular}
\end{ruledtabular}
\end{table*}

\begin{table}[t]
\caption{One-body moments and transition moments in $\muN$ in the constituent quark model, for the charge states
and for the $I=\tfrac12$ average. $\mu_{qq}$ is the moment of the light diquark and $\mu_q$ that of the light
quark of the meson. For $\Sc^{(*)}\to\Lc$ the last column is the overlap with $|\Db^{(*)0}\Lc^+\rangle$. The
lattice values are those of Table~III of Ref.~\cite{Can:2014} for the linear and the quadratic extrapolation in
$m_\pi^2$, with $\mu_{\Sc^+}=\tfrac12(\mu_{\Sc^{++}}+\mu_{\Sc^0})$ in the average.}
\label{tab:moments}
\begin{ruledtabular}
\footnotesize
\setlength{\tabcolsep}{1.5pt}
\begin{tabular}{llcccc}
 & expression & $++$ & $+$ & $0$ & $I=\tfrac12$\\
\hline
$\mu_{\Sc}$ & $\tfrac23\mu_{qq}-\tfrac13\mu_c$ & $2.343$ & $0.482$ & $-1.380$ & $1.723$\\
$\mu_{\Sc}$ (lat.) & linear & $1.50(20)$ & & $-0.88(10)$ & $1.10$\\
 & quadratic & $2.03(39)$ & & $-1.12(20)$ & $1.50$\\
$\mu_{\Scs}$ & $\mu_{qq}+\mu_c$ & $4.140$ & $1.348$ & $-1.445$ & $3.209$\\
$\mu_{\Sc\Scs}$ & $-\tfrac{\sqrt2}3(\mu_{qq}-2\mu_c)$ & $-1.362$ & $-0.046$ & $1.271$ & $-0.923$\\
$\mu_{\Sc\Lc}$ & $(\mu_u-\mu_d)/\sqrt3$ & & $1.612$ & & $-0.931$\\
$\mu_{\Scs\Lc}$ & $\sqrt{2/3}(\mu_u-\mu_d)$ & & $2.280$ & & $-1.316$\\
$\mu_{\Lc^+}$ & $\mu_c$ & \multicolumn{3}{c}{$0.417$} & $0.417$\\
\hline
 & & $\Db^{(*)0}$ & $D^{(*)-}$ & & \\
\hline
$\mu_{\Dbs}$ & $\mu_q-\mu_c$ & $1.445$ & $-1.348$ & & $-0.417$\\
$\mu_{\Db\Dbs}$ & $-(\mu_q+\mu_c)$ & $-2.279$ & $0.514$ & & $-0.417$\\
\hline
$\mu_p$, $\mu_n$ & experiment & \multicolumn{3}{c}{$2.793$, $-1.913$} & \\
$\mu_{\etac\Jpsi}$ & $-2\mu_c$ & \multicolumn{3}{c}{$-0.834$} & \\
\end{tabular}
\end{ruledtabular}
\end{table}

The pole positions obtained in the two parameter sets are listed in Table~\ref{tab:poles}. The poles of set I
agree with those of Ref.~\cite{Xiao:2019aya} within $0.2\MeV$ in the mass and $0.1\MeV$ in the width, except for
the $\Db\Sc$ state. For this state, the pole position reported in Ref.~\cite{Xiao:2019aya}, $4306.4-7.6i\MeV$, is
reproduced if the weakly coupled $\Dbs\Lc$ channel is taken on its first Riemann sheet. On the sheet specified
above, we find the pole at $4306.6-7.0i\MeV$. In set II, the masses of the three observed states are reproduced by construction, and their widths are $10.1$, $23.7$ and $3.0\MeV$, compared with the
experimental values $9.8^{+4.6}_{-5.2}$, $20.6^{+10.0}_{-11.2}$ and $6.4^{+6.0}_{-2.8}\MeV$. The model also predicts
four partner states, a $\Db\Scs$ state at $4378\MeV$ and three $\Dbs\Scs$ states at $4505$, $4517$ and
$4525\MeV$. Since the constituents are treated as stable particles and the width of the $\Scs$ is comparable
with or larger than the binding energies of these four states, their pole positions should be regarded as
estimates (see Sec.~\ref{sec:measured}).

We adopt the spin assignment of Ref.~\cite{Xiao:2019aya}, $\tfrac12^-$ for $P_c(4440)$ and $\tfrac32^-$ for
$P_c(4457)$, which we refer to as ordering A. Within the present kernel, this assignment is favored by the
widths. Since the $\Jpsi N$ channel couples to the $\Dbs\Sc$ channel with the weights $\tfrac56$ for
$J=\tfrac12$ and $\tfrac13$ for $J=\tfrac32$, the $\tfrac12^-$ state is broader than the $\tfrac32^-$ state. For
the opposite ordering B, the observed masses require $r=3.5$--$4.4$, the resulting widths show the reverse of the
observed pattern, and the $\Db\Sc$ state lies $15\MeV$ below $P_c(4312)$ mass. This argument, however, does not
apply to models with one-pion exchange, in which the tensor force modifies the relation between spin and
width~\cite{Du:2020,Du:2021,Yamaguchi:2020}. We therefore present the magnetic moments for both orderings in
Sec.~\ref{sec:moments}.

The compositeness of channel $l$ and the energy-dependence term are defined
as~\cite{Gamermann:2009uq,Hyodo:2011qc,Aceti:2012dd,Sekihara:2014kya}
\begin{equation}
\begin{split}
X_l&=-g_l^2\,\frac{\dd G_l}{\dd\rs}\Big|_{\zR},\qquad \sum_lX_l+Z=1 ,\\
Z&=-\sum_{l',l}g_{l'}G_{l'}\,\frac{\partial V_{l'l}}{\partial\rs}\,G_l\,g_l\Big|_{\zR}.
\end{split}
\label{eq:sumrule}
\end{equation}
For a resonance, the $X_l$ are complex and cannot be interpreted as probabilities. All states are dominated by a
single channel, with $|X_{\rm main}-1|<0.03$ and $|X_l|<0.02$ for the other channels.

\subsection{Photon couplings}
\label{subsec:EMint}

The photon couples both to the constituents and to their interaction. Each coupling enters through the Feynman
rule $-ie\,\epsilon_\mu\Gt^\mu$, where the vertex function $\Gt^\mu$ is the matrix element of the current in units
of $e$. The electromagnetic current is thus the sum of one-body currents and a photon--four-hadron vertex.

\subsubsection{One-body currents}
\label{subsubsec:onebody}

The charge couplings are given by the convection currents $Q_M(k+k')^\mu$ of the mesons, which are diagonal in
the polarization of the $\Dbs$, and $Q_B(p+p')^\mu/(2M_B)$ of the baryons in the nonrelativistic
reduction~\cite{Sekihara:2010uz}. The electric charges of the meson and the baryon are
$(Q_M,Q_B)=(-\tfrac23,\tfrac53)$ in the $\Db^{(*)}\Sc^{(*)}$ channels, obtained as the isospin average over the
two charge states, and $(0,1)$ in the $\Db^{(*)}\Lc$ and hidden-charm channels. In all channels, the
baryon-number and anticharm probes have the charges $(0,1)$ and $(1,0)$, respectively, while the charm probe has
$(-1,1)$ in the open-charm channels and $(0,0)$ in the hidden-charm channels.

The magnetic moments and the M1 transition moments of the hadrons are taken from the constituent quark model,
which is consistent with the HQSS structure of the fields $\Hb$, $S^i$ and $\Psi$. We use $\mu_u=1.862$,
$\mu_d=-0.931$ and $\mu_c=0.417\,\muN$ ($m_{u,d}=336\MeV$, $m_c=1.50\GeV$) together with the experimental nucleon
moments. The resulting values are listed in Table~\ref{tab:moments}, together with the lattice QCD results of
Ref.~\cite{Can:2014} for the $\Sc$. The magnetic current of a hadron is given by
$\bm J_{\rm mag}(\vq)=i(\bm\mu_{\rm op}/e)\times\vq$. In the channel basis, the one-body magnetic operator is
reduced by means of the Wigner--Eckart theorem,
\begin{equation}
\langle l';JM'|\hat{\bm\mu}^{(1)}|l;JM\rangle=\mathcal M^{(1)}_{l'l}(J)\,\frac{\langle JM'|\vJ|JM\rangle}{J}.
\label{eq:WE}
\end{equation}
The matrix $\mathcal M^{(1)}=\mathcal M^{\rm M}+\mathcal M^{\rm B}$ consists of a meson-line and a baryon-line
contribution (see Appendix~\ref{app:moments}). Its diagonal elements are the additive moments of the two-hadron
states, and its off-diagonal elements describe transitions between channels with a common spectator. With the
isospin weights $\tfrac23$ and $\tfrac13$, the light-quark contribution to the moment of the $\Dbs$ in the
$\Db^{(*)}\Sc^{(*)}$ channels is $\tfrac13(\mu_u+2\mu_d)$. This contribution vanishes for $\mu_u=-2\mu_d$, which
yields $\langle\mu_{\Dbs}\rangle=-\mu_c$.

\subsubsection{Photon--four-hadron vertex}
\label{subsubsec:gauged}

The coupling of the photon to the interaction is described by the photon--four-hadron vertex
\begin{equation}
\Gt^\mu_{l'l}(P,P')=\Gt^\mu_{{\rm min},l'l}+\Gt^\mu_{M,l'l}+\Gt^\mu_{E,l'l},
\label{eq:Gamma}
\end{equation}
where $P$ and $P'$ are the total four-momenta before and after the absorption of the photon. The first term is
determined by the kernel through current conservation. The magnetic and electric contact terms,
$\Gt^\mu_{\rm tr}=\Gt^\mu_M+\Gt^\mu_E$, are transverse and contain couplings that are not fixed by the model. To
the best of our knowledge, this vertex has not been derived previously for the $\Db^{(*)}\Sc^{(*)}$
coupled-channel system, and its construction is one of the new elements of the present work.

\emph{Gauged kernel.} An energy-dependent contact interaction corresponds to a Lagrangian with time derivatives.
Gauging these derivatives generates the part $\Gt^\mu_{\rm min}$ of the vertex, which is constrained by current
conservation, $q_\mu\Gt^\mu_{l'l}=-Q_H[V_{l'l}(\rsp)-V_{l'l}(\rs)]$, where $Q_H$ is the charge of the
state~\cite{Kvinikhidze:1999,Borasoy:2005zg}. For an arbitrary energy dependence of the kernel, a solution is given
by
\begin{equation}
\begin{split}
\Gt^\mu_{{\rm min},l'l}(P,P')&=-Q_H\,\frac{(P+P')^\mu}{s'-s}\Big[V_{l'l}(\rsp)-V_{l'l}(\rs)\Big]\\
&\xrightarrow{\,s'\to s\,}\ -Q_H\,\frac{(P+P')^\mu}{2\rs}\,\frac{\partial V_{l'l}}{\partial\rs},
\end{split}
\label{eq:Gmin}
\end{equation}
which is unique up to transverse terms. For the kernel of Eq.~\eqref{eq:HGvalues}, the derivative is obtained
from
\begin{equation}
\frac{\partial k^0_l}{\partial\rs}=\frac{\omega_l}{\rs},\qquad
\frac{\partial p^2_{D^*,l'l}}{\partial\rs}=-\frac{2\big(k^0_{l'}\omega_l+\omega_{l'}k^0_l\big)}{\rs},
\label{eq:dV}
\end{equation}
where $\omega_l=(s+M_l^2-m_l^2)/(2\rs)$ is the baryon energy. The derivative of the $D^*$-exchange factor in
$\mu_{12}$ must be included to satisfy the Ward identity of Sec.~\ref{subsec:photon}. For the
Weinberg--Tomozawa kernel, Eq.~\eqref{eq:Gmin} reduces to the vertex of Ref.~\cite{Sekihara:2010uz}. This vertex
vanishes for a neutral state but contributes for the charged $P_c^+$. In the Breit frame, it contributes only to
$\GE$.

\emph{Magnetic contact vertex.} Gauge invariance does not determine the transverse part of the vertex.
Transverse vertices are constructed from the field strength $F_{\mu\nu}$, and each of them introduces a new
coupling. We consider the operators that contain one magnetic field and no further derivative,
$\delta\mathcal L_M=e\,\vB\cdot\bm{\mathcal X}$, where $\bm{\mathcal X}$ is a two-hadron operator that transforms
as a vector in the light-quark spins. In the present model space, parity, HQSS, time-reversal invariance,
Hermiticity and the isospin structure of the photon allow $34$ such operators (see Appendix~\ref{app:fivepoint}).
For an $I=\tfrac12$ state, they contribute through ten combinations $\bar c_s$ of their couplings,
$\bm{\mathcal X}=\sum_s\bar c_s\bm{\mathcal X}^{(s)}$. In the $\Db^{(*)}\Sc^{(*)}$ channels, these operators act
on the spin $\vs_q$ of the light quark in the meson and on the spin $\bm S_{qq}$ of the light diquark in the
baryon, $\bm{\mathcal X}^{(a)}=2\vs_q$, $\bm{\mathcal X}^{(b)}=2\bm S_{qq}$ and
$\mathcal X^{(c)j}=-2\sigma_q^iQ^{ij}_{qq}$, where $Q^{ij}_{qq}$ is the quadrupole operator of the diquark, see Eq.~\eqref{eq:Xquark}. For an incoming photon, $\vB\to i\vq\times\bm\epsilon$, and the vertex takes the form
\begin{equation}
\Gt^\mu_{M,l'l}=2\sqrt{m_{l'}m_l}\;i\,\epsilon^{\mu\nu\rho\sigma}v_\nu q_\rho\,\mathcal X_{\sigma,l'l},
\label{eq:GM}
\end{equation}
where $v^\mu=(P+P')^\mu/[(P+P')^2]^{1/2}$, $\epsilon^{0123}=+1$, and $\mathcal X^\mu$ is the spin four-vector,
which reduces to $(0,\bm{\mathcal X})$ in the rest frame of the meson--baryon pair. The factor
$2\sqrt{m_{l'}m_l}$ converts the nonrelativistic normalization of the meson fields in the operators to the
normalization of Eq.~\eqref{eq:Gloop}. The vertex is transverse, $q_\mu\Gt^\mu_M=0$, and in the Breit frame at
$s=s'$ its components are $\Gt^0_M=0$ and $\bm\Gt_M=2\sqrt{m_{l'}m_l}\,i\,\bm{\mathcal X}\times\vq$. Between
states of total spin $J$, the Wigner--Eckart theorem yields
\begin{equation}
\begin{split}
\langle l';JM'|\bm\Gt_M|l;JM\rangle&=2\sqrt{m_{l'}m_l}\,i\mathcal A^{(J)}_{l'l}(\vJ\times\vq)_{M'M},\\
\mathcal A^{(J)}_{l'l}&=\sum_s\bar c_s\,\Xis^{(s)}_{l'l}(J),
\end{split}
\label{eq:Jtr}
\end{equation}
where the spin factors $\Xis^{(s)}_{l'l}(J)=\langle l';JJ|\mathcal X^{(s)}_z|l;JJ\rangle/J$ are listed in
Table~\ref{tab:gamma}.

\emph{Electric contact vertex.} Parity forbids operators that contain an electric field without a further
derivative. The lowest electric operators are therefore
\begin{equation}
\delta\mathcal L_E=e\,d\,(\bm\nabla\cdot\vE)\,\mathcal X ,
\label{eq:LE}
\end{equation}
where $\mathcal X$ is a two-hadron operator that is a scalar in the light-quark spins. Since
$\bm\nabla\cdot\vE=v_\nu\partial_\mu F^{\mu\nu}$ in the rest frame of the pair, the corresponding vertex is
\begin{equation}
\Gt^\mu_{E,l'l}=2\sqrt{m_{l'}m_l}\,\Ecal_{l'l}\big[q^2v^\mu-q^\mu\,(q\cdot v)\big],
\label{eq:GE}
\end{equation}
where $\Ecal$ is a real symmetric matrix of couplings with dimension $E^{-5}$. This vertex is transverse, and in
the Breit frame at $s=s'$ its only nonvanishing component is $\Gt^0_E=-Q^2\,2\sqrt{m_{l'}m_l}\,\Ecal_{l'l}$. It
therefore changes the slope of $\GE$ without changing the charge. In the $\Db^{(*)}\Sc^{(*)}$ channels, the scalar
operators are $1$ and $\vs_q\cdot\bm S_{qq}$, the latter with the eigenvalues $-2$ and $+1$ for light-quark spin
$\tfrac12$ and $\tfrac32$, respectively. Consequently, the matrix $\Ecal$ has the same HQSS structure as the
kernel, with one coupling for each light-quark spin. Including the other pairs of channel types, there are $22$
such operators, which contribute to an $I=\tfrac12$ state through seven combinations (see
Appendix~\ref{app:fivepoint}).

The magnetic contact operators are analogous to the four-nucleon--one-photon counterterms in the pionless
effective field theory of the nucleon--nucleon interaction~\cite{Chen:1999tn}. In the present case, however,
they are constrained by HQSS and connect channels with different heavy-meson and heavy-baryon spins.

\subsubsection{Contact couplings}
\label{subsubsec:contact}

The contact couplings are neither fixed by the kernel nor constrained by experimental data. We set their
central values to zero for the following three reasons. First, neither their magnitudes nor their signs are
known, and the value zero favors neither sign and lies at the center of the variation considered below. Second,
with this choice the central form factors are determined solely by the one-body currents and the gauged kernel,
which are fixed by the strong-interaction model, the charges and the constituent moments. Third, only an
order-of-magnitude estimate of these couplings is available. The anomalous $\pi\rho\gamma$ and
$\pi\omega\gamma$ exchange currents yield magnetic couplings of about $10\GeV^{-4}$ in magnitude with
undetermined signs, which change the magnetic moments by at most $0.07\,\muN$ (see Appendix~\ref{app:fivepoint}).
This change is smaller than the variation adopted in this work.

The range of the magnetic couplings is estimated by dimensional analysis, $|\bar c|\sim|V|/\Lambda^2$. With the
diagonal kernel of the dominant channels, $|V|=16$ to $29\GeV^{-2}$ in the normalization of the operators, and
$\Lambda=0.7$ to $1\GeV$, this estimate gives values between $15$ and $60\GeV^{-4}$. We therefore adopt the bounds
\begin{equation}
|\bar c_s|\leq C_s,\qquad C_s=\begin{cases}30\GeV^{-4}&s=a,b,c,d,f,f',\\ 6\GeV^{-4}&s=h,h',k,n,\end{cases}
\label{eq:domain}
\end{equation}
where the first six combinations act between the $\Db^{(*)}\Sc^{(*)}$ and $\Db^{(*)}\Lc$ channels and the last
four involve the hidden-charm channels. For the transitions to the hidden-charm channels, the size is reduced by
the factor $m_\rho^2/|p_{D^*}^2-m_{D^*}^2|\simeq0.2$ of the $D^*$ exchange, and the same bound is used for the
diagonal combination $\bar n$ of these channels. Since the contact terms are linear in the couplings,
$\GM(0)=\GM(0)|_{\bar c=0}+\sum_sK_s\bar c_s$ with complex coefficients $K_s$. The maximal shifts of the real
part are
\begin{equation}
\delta_{\rm box}=\sum_s|\Rl K_s|\,C_s,\qquad
\delta_{\rm ell}=\Big[\sum_s(\Rl K_s)^2C_s^2\Big]^{1/2}
\label{eq:box}
\end{equation}
for independently bounded couplings and for couplings inside the ellipsoid $\sum_s(\bar c_s/C_s)^2\leq1$,
respectively. The ranges of the magnetic moments in Sec.~\ref{sec:moments} are obtained with $\delta_{\rm box}$.

The contribution of the contact terms is $\psi^{\sf T}\tilde{\mathcal A}\,\psi$ with $\psi_l=g_lG_l(\zR)$ and
$\tilde{\mathcal A}_{l'l}=2\sqrt{m_{l'}m_l}\,\mathcal A_{l'l}$, where $G_l(\zR)$ depends on the subtraction
constants. A shift $G\to G+C$ by a constant diagonal matrix $C$, compensated in the kernel by
$V^{-1}\to V^{-1}+C$, leaves $T$, the couplings $g_l$, the one-body loops and $Z$ unchanged. The contact term,
however, remains unchanged only if
\begin{equation}
\tilde{\mathcal A}\to(\one+VC)^{-1}\,\tilde{\mathcal A}\,(\one+CV)^{-1}\quad\text{at }\rs=\zR .
\label{eq:matching}
\end{equation}
Hence, the couplings $\bar c_s$ and the bounds \eqref{eq:domain} refer to the subtraction constants of
Eq.~\eqref{eq:Gloop}, and only the complete current is independent of the regularization scheme.

The electric couplings are not determined by the model either, and they are not included in the form factors.
Inserted between the amplitudes $\psi_l$, the vertex \eqref{eq:GE} shifts the charge radius by
\begin{equation}
\delta\MSR_E=12\sum_{l',l}\psi_{l'}\sqrt{m_{l'}m_l}\,\Ecal_{l'l}\,\psi_l ,
\label{eq:drE}
\end{equation}
whose size is estimated in Sec.~\ref{subsec:constituents}.

\addedprev{Two further properties support the choice of zero central values. First, the magnetic moments and the charge
radii are linear in the contact couplings. Their values at $\bar c_s=0$ and $\Ecal=0$ are therefore the
expectation values for any distribution of the couplings that is symmetric under a change of their sign. Second,
the electric and magnetic contact vertices enter the amplitude in the same way, so that under a change of the
subtraction constants the electric couplings transform as $\tilde{\mathcal A}$ does in Eq.~\eqref{eq:matching}. This
transformation is linear and homogeneous, and zero is the only value of the couplings that is the same for all
subtraction constants related in this way. The central values are therefore independent of the subtraction
constants within this class, whereas a nonzero central value would refer to one particular choice.}

\subsection{Form factors at the pole}
\label{subsec:photon}

\begin{figure*}[t]
\centering
\begin{tikzpicture}[scale=0.85, every node/.style={transform shape},
  meson/.style={dashed,thick},
  baryon/.style={thick},
  photon/.style={decorate,decoration={snake,amplitude=1.2pt,segment length=5pt},thick},
  blob/.style={ellipse,draw,fill=gray!35,minimum width=0.95cm,minimum height=0.7cm,inner sep=0pt},
  mom/.style={-{Stealth[length=4.5pt,width=3.5pt]},semithick}]
\begin{scope}[xshift=0cm]
  \node[blob] (TL) at (0,0) {$T$};
  \node[blob] (TR) at (3.0,0) {$T$};
  \draw[meson] (TL.north east) .. controls (0.9,0.9) and (2.1,0.9) .. (TR.north west);
  \draw[baryon] (TL.south east) .. controls (0.9,-0.9) and (2.1,-0.9) .. (TR.south west);
  \draw[photon] (1.5,0.68) -- (1.5,1.9);
  \draw[meson] (-1.0,0.45) -- (TL.north west);
  \draw[baryon] (-1.0,-0.45) -- (TL.south west);
  \draw[meson] (TR.north east) -- (4.0,0.45);
  \draw[baryon] (TR.south east) -- (4.0,-0.45);
  \node at (1.5,-1.5) {$T^\mu_{\gamma1}$};
  \node[anchor=east] at (-1.05,0.45) {\footnotesize $M_l$};
  \node[anchor=east] at (-1.05,-0.45) {\footnotesize $B_l$};
  \node[anchor=west] at (4.05,0.45) {\footnotesize $M_{l'}$};
  \node[anchor=west] at (4.05,-0.45) {\footnotesize $B_{l'}$};
  \node[anchor=west] at (1.6,1.75) {\footnotesize $\gamma^*(q)$};
  \draw[mom] (1.28,1.80) -- (1.28,1.10);
  \draw[mom] (-0.95,-0.80) -- (-0.35,-0.80);
  \node[anchor=north] at (-0.65,-0.82) {\footnotesize $P$};
  \draw[mom] (3.35,-0.80) -- (3.95,-0.80);
  \node[anchor=north] at (3.65,-0.82) {\footnotesize $P'=P+q$};
\end{scope}
\begin{scope}[xshift=6.9cm]
  \node[blob] (TL) at (0,0) {$T$};
  \node[blob] (TR) at (3.0,0) {$T$};
  \draw[meson] (TL.north east) .. controls (0.9,0.9) and (2.1,0.9) .. (TR.north west);
  \draw[baryon] (TL.south east) .. controls (0.9,-0.9) and (2.1,-0.9) .. (TR.south west);
  \draw[photon] (1.5,-0.68) -- (1.5,-1.9);
  \draw[meson] (-1.0,0.45) -- (TL.north west);
  \draw[baryon] (-1.0,-0.45) -- (TL.south west);
  \draw[meson] (TR.north east) -- (4.0,0.45);
  \draw[baryon] (TR.south east) -- (4.0,-0.45);
  \node at (1.5,1.5) {$T^\mu_{\gamma2}$};
  \node[anchor=west] at (1.6,-1.75) {\footnotesize $\gamma^*(q)$};
  \draw[mom] (1.28,-1.80) -- (1.28,-1.10);
  \draw[mom] (-0.95,0.80) -- (-0.35,0.80);
  \node[anchor=south] at (-0.65,0.82) {\footnotesize $P$};
  \draw[mom] (3.35,0.80) -- (3.95,0.80);
  \node[anchor=south] at (3.65,0.82) {\footnotesize $P'$};
\end{scope}
\begin{scope}[xshift=13.3cm]
  \node[blob] (TL) at (0,0) {$T$};
  \node[blob] (TR) at (3.8,0) {$T$};
  \node[circle,draw,fill=black,inner sep=1.6pt] (V) at (1.9,0) {};
  \draw[meson] (TL.north east) .. controls (0.7,0.8) and (1.4,0.55) .. (V);
  \draw[baryon] (TL.south east) .. controls (0.7,-0.8) and (1.4,-0.55) .. (V);
  \draw[meson] (V) .. controls (2.4,0.55) and (3.1,0.8) .. (TR.north west);
  \draw[baryon] (V) .. controls (2.4,-0.55) and (3.1,-0.8) .. (TR.south west);
  \draw[photon] (V) -- (1.9,1.9);
  \draw[meson] (-1.0,0.45) -- (TL.north west);
  \draw[baryon] (-1.0,-0.45) -- (TL.south west);
  \draw[meson] (TR.north east) -- (4.8,0.45);
  \draw[baryon] (TR.south east) -- (4.8,-0.45);
  \node at (1.9,-1.9) {$T^\mu_{\gamma3}$};
  \node[anchor=west] at (2.0,1.75) {\footnotesize $\gamma^*(q)$};
  \draw[mom] (1.68,1.80) -- (1.68,1.10);
  \draw[mom] (-0.95,-0.80) -- (-0.35,-0.80);
  \node[anchor=north] at (-0.65,-0.82) {\footnotesize $P$};
  \draw[mom] (4.15,-0.80) -- (4.75,-0.80);
  \node[anchor=north] at (4.45,-0.82) {\footnotesize $P'$};
  \node[anchor=north] at (1.9,-0.85) {\footnotesize $\Gamma^\mu_{\rm min}+\Gamma^\mu_{\rm tr}$};
\end{scope}
\end{tikzpicture}
\caption{The three classes of diagrams of the photon-coupled amplitude $T^\mu_\gamma$ with a double pole of the
state: photon attached to the meson line ($T_{\gamma1}$), to the baryon line ($T_{\gamma2}$) and to the
four-hadron vertex ($T_{\gamma3}$). The shaded ellipses are the amplitudes $T(\rs)$ (left) and $T(\rsp)$
(right), dashed (solid) lines are mesons (baryons), and the wavy line is the incoming virtual photon or another
probe current. $P$ and $P'=P+q$ are the total momenta before and after the absorption of $q$, with $s=P^2$ and
$s'=P'^2$. The loops of $T_{\gamma1,2}$ are matrices in channel space, because the magnetic transitions
$\Dbs\to\Db\gamma$, $\Scs\to\Sc\gamma$ and $\Sc^{(*)}\to\Lc\gamma$ connect different channels. The vertex of
$T_{\gamma3}$ is the sum of the gauged kernel $\Gamma_{\rm min}$ and the transverse contact terms
$\Gamma_{\rm tr}$.}
\label{fig:diagrams}
\end{figure*}
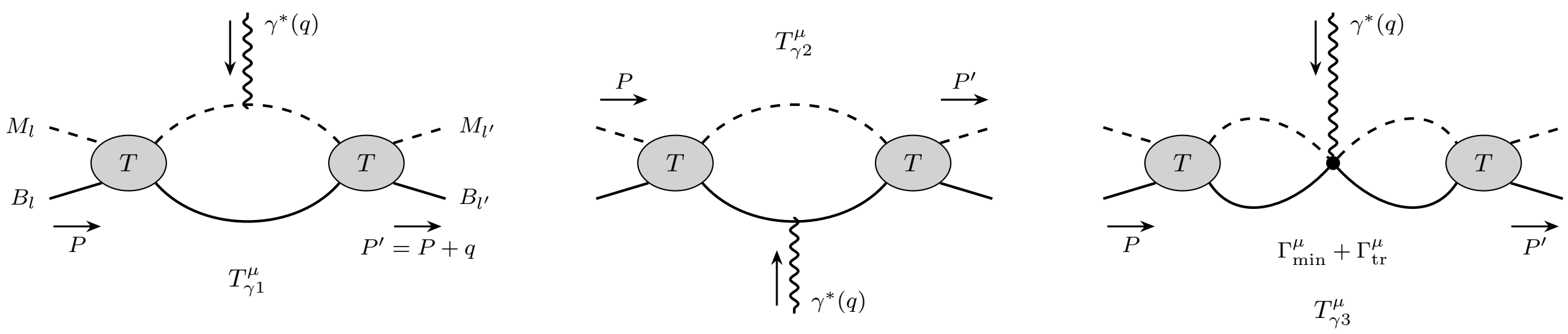

Attaching the photon to the scattering series yields three classes of diagrams with a double pole, as shown in
Fig.~\ref{fig:diagrams}. Diagrams in which the photon couples to an external leg have only a single pole and
therefore do not contribute to Eq.~\eqref{eq:residue}~\cite{Sekihara:2008qk,Borasoy:2005zg}. In channel space,
the photon-coupled amplitude reads
\begin{equation}
\begin{split}
T^\mu_\gamma(\rsp,\rs)&=T(\rsp)\,\Pi^\mu(P,P')\,T(\rs),\\
\Pi^\mu_{l'l}&=D^\mu_{M;l'l}+D^\mu_{B;l'l}+G_{l'}(\rsp)\,\Gt^\mu_{l'l}\,G_l(\rs),
\end{split}
\label{eq:Xdef}
\end{equation}
where $D^\mu_M$ and $D^\mu_B$ denote the triangle loops with the photon attached to the meson and baryon lines,
respectively~\cite{Sekihara:2010uz}, and $\Gt^\mu$ is the vertex of Eq.~\eqref{eq:Gamma}. Inserting the pole terms
of $T$ and comparing with Eq.~\eqref{eq:residue}, we obtain
\begin{equation}
\langle\Pc(P')|J^\mu|\Pc(P)\rangle=\sum_{l',l}g_{l'}\,\Pi^\mu_{l'l}(P,P')\,g_l\Big|_{\rs=\rsp=\zR} ,
\label{eq:master}
\end{equation}
where the loops are continued to the Riemann sheet of the pole. Note that the couplings enter without complex
conjugation. At the pole, the Breit frame is defined by $P=(E,-\vq/2)$ and $P'=(E,\vq/2)$ with
$E=\sqrt{\zR^2+Q^2/4}$. The form factor $F_P$ of a probe with the charges $(Q^P_{M,l},Q^P_{B,l})$ and the total
charge $Q^P_H$, together with the magnetic form factor, is given by
\begin{equation}
\begin{aligned}
F_P&=\sum_lg_l^2\Big[Q^P_{M,l}\,\EM_l+Q^P_{B,l}\,\EB_l\Big]+Q^P_H\,\frac{E}{\zR}\,Z ,\\
\GM&=\sum_{l',l}g_{l'}\Big[\mathcal M^{\rm M}_{l'l}\,\MM_{l'l}+\mathcal M^{\rm B}_{l'l}\,\MB_{l'l}\Big]g_l\\
&\quad+2M_pJ\sum_{l',l}\psi_{l'}\,2\sqrt{m_{l'}m_l}\,\mathcal A^{(J)}_{l'l}\,\psi_l ,
\end{aligned}
\label{eq:final}
\end{equation}
where $\mathbb E^{\rm M,B}_l(Q^2)$ and $\mathbb M^{\rm M,B}_{l'l}(Q^2)$ are the electric and magnetic loop functions
of Appendix~\ref{app:loops}, and $\GE=F_P$ for the electric charges. The matrices $\mathcal M^{\rm M,B}$ are given
in units of $\muN$. The term proportional to $Z$ represents the contribution of the gauged kernel
\eqref{eq:Gmin}. In the nonrelativistic limit, the meson and the baryon are located at the fractions
\begin{equation}
\xi_M=\frac{M_l}{m_l+M_l},\qquad \xi_B=\frac{m_l}{m_l+M_l}
\label{eq:xi}
\end{equation}
of the relative distance from the center of mass.

The one-body loops satisfy the Ward identity $q_\mu D^\mu_{M,B;ll}=Q_{M,B;l}[G_l(\rs)-G_l(\rsp)]$ on every Riemann
sheet. Combining this identity with $q_\mu\Gt^\mu_{\rm min}$ and the scattering equation, we obtain
\begin{equation}
\begin{split}
q_\mu T^\mu_\gamma(\rsp,\rs)=-Q_H\big[&T(\rsp)G(\rsp)V(\rs)\\
&-V(\rsp)G(\rs)T(\rs)\big],
\end{split}
\label{eq:WTfull}
\end{equation}
which contains only single poles. The coefficient of the double pole is therefore conserved. At $Q^2=0$, where
$\EM_l(0)=\EB_l(0)=-\dd G_l/\dd\rs$, this leads to
\begin{equation}
\GE(0)=Q_H\Big[\sum_lX_l+Z\Big]=Q_H .
\label{eq:GE0}
\end{equation}
The charge normalization is thus equivalent to the compositeness sum rule \eqref{eq:sumrule}, in which the
gauged kernel provides the term $Z$. Numerically, Eq.~\eqref{eq:GE0} is satisfied to better than $10^{-5}$ at all
poles. This result extends the relation between the normalization of the form factor and the compositeness
established in Refs.~\cite{Sekihara:2010uz,Sekihara:2014kya} to a charged state. While this identity fixes the
charge, it imposes no constraint on the contact currents of Sec.~\ref{subsubsec:contact} or on the $Q^2$ dependence of the form factors.

The residue is evaluated on the Riemann sheet of the pole. For an open channel, the photon loops on the second
sheet are obtained by adding the discontinuity across the unitarity cut (see Appendix~\ref{app:loops}). In
addition, these loops have a triangle singularity at spacelike momentum transfer. We define the form factor as
the branch of the residue that is analytic in $Q^2$ and contains the point $Q^2=0$ (see Sec.~\ref{sec:open}).

\section{Form factors and radii}
\label{sec:poles}

\subsection{Form factors}
\label{subsec:ff}

\begin{figure*}[t]
\centering
\includegraphics[width=0.93\textwidth]{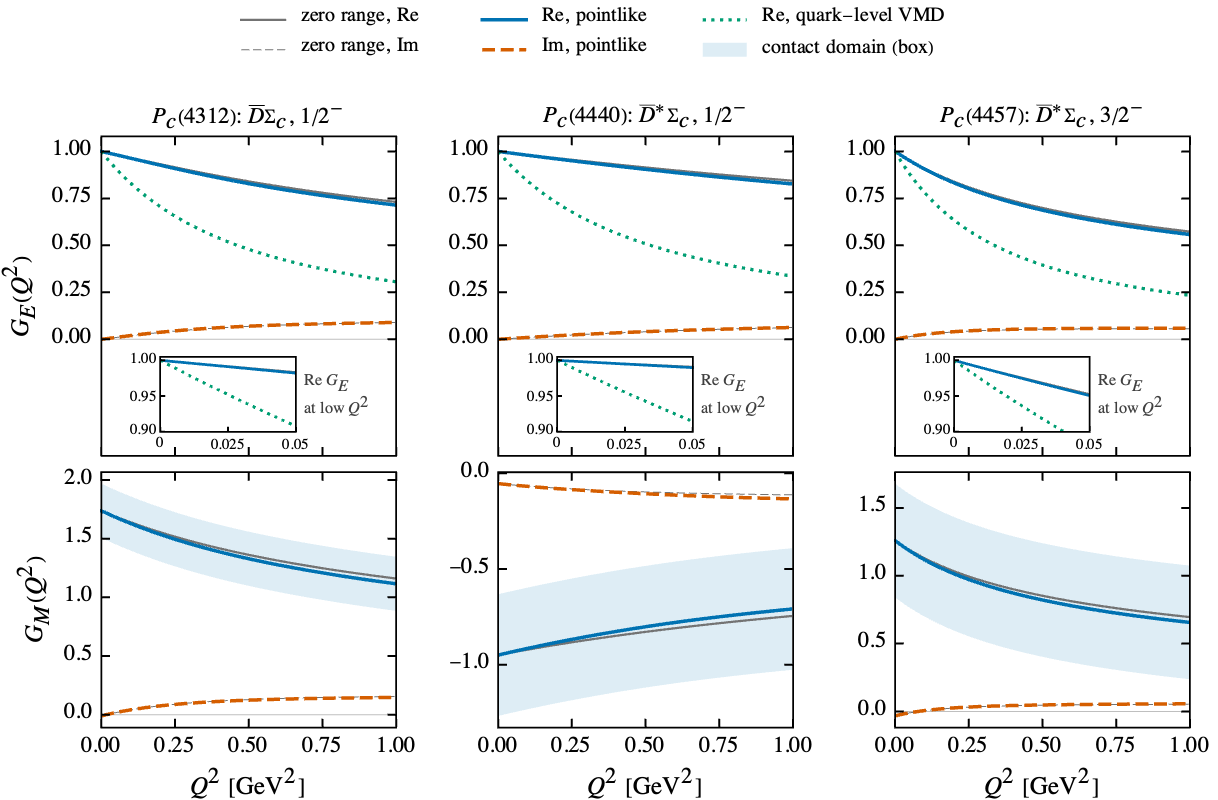}
\caption{Electric (top) and magnetic (bottom) form factors of the three observed states at the poles of set II
for pointlike constituents. Thick solid (dashed) lines are real (imaginary) parts. Thin gray lines are the
zero-range form of Eq.~\eqref{eq:ZRff}. Dotted lines are the real part of $\GE$ with quark-level vector-meson
dominance for the constituents (Sec.~\ref{subsec:constituents}). The insets show the region of low $Q^2$. The
bands in the lower panels are ${\rm Re}\,G_M\pm\delta_{\rm box}$, where $\delta_{\rm box}$ is the maximal shift from of the magnetic contact couplings,
Eq.~\eqref{eq:box}, around $\Rl\GM$.}
\label{fig:ff}
\end{figure*}

In Fig.~\ref{fig:ff}, we show the form factors of the three observed states in set II for $Q^2\leq1\GeV^2$. At
$Q^2=1\GeV^2$, the electric form factors are $0.71+0.09i$, $0.83+0.06i$ and $0.56+0.06i$ for $P_c(4312)$,
$P_c(4440)$ and $P_c(4457)$, respectively. The form factor of the most weakly bound state, $P_c(4457)$, decreases
most rapidly. The imaginary parts of $\GE$ vanish at $Q^2=0$ and remain below $0.10$.

The $Q^2$ dependence is characteristic of a weakly bound two-body system. We denote by $m$ and $M$ the meson and
baryon masses of the dominant channel, by $\mured=mM/(m+M)$ its reduced mass, and by
\begin{equation}
B_c=m+M-\zR=B+\frac i2\Gamma_R,\qquad \gamma=\sqrt{2\mured B_c},
\label{eq:Bc}
\end{equation}
the complex binding energy and the binding momentum. For a zero-range wave function $\psi\propto e^{-\gamma x}/x$,
the form factor of a constituent located at the fraction $\xi$ of the relative distance $x$ from the center of
mass is
\begin{equation}
F_{\rm ZR}(Q;\xi)=\frac{2\gamma}{\xi Q}\arctan\frac{\xi Q}{2\gamma}.
\label{eq:ZRff}
\end{equation}
The thin lines in Fig.~\ref{fig:ff} show $\tfrac53F_{\rm ZR}(Q;\xi_B)-\tfrac23F_{\rm ZR}(Q;\xi_M)$ for $\GE$ and the
corresponding combination of the baryon-line and meson-line contributions to $\GM(0)$ for $\GM$. These curves
contain no parameter other than the pole position. The full loop results agree with them within $0.02$ for $\GE$
and within $0.05$ for $\GM$ up to $Q^2=1\GeV^2$. For pointlike constituents, the form factors are therefore
essentially determined by the complex binding energy, the masses and charges of the constituents and the static
moments.

\subsection{Mean squared distance and the zero-range relation}
\label{subsec:radii}

\begin{table*}[t]
\caption{Complex mean squared distance $\MSD$ between the constituents of the dominant channel,
Eq.~\eqref{eq:MSD}, and its zero-range value $1/(4\mured B_c)$ in the two parameter sets, and mean squared radii
of the charge, of the magnetization and of the charm current in set II. All values are in fm$^2$, for pointlike
constituents and without contact currents. The charm radius is not normalized, because $\FC(0)=0$.}
\label{tab:radii}
\begin{ruledtabular}
\begin{tabular}{lccccccc}
 & \multicolumn{2}{c}{set I} & \multicolumn{2}{c}{set II} & \multicolumn{3}{c}{radii, set II}\\
state & $\MSD$ & $1/(4\mured B_c)$ & $\MSD$ & $1/(4\mured B_c)$ & $\MSR_E$ & $\MSR_M$ & $\MSR_C$\\
\hline
$\Db\Sc(\tfrac12^-)$ & $0.59-0.26i$ & $0.52-0.26i$ & $0.85-0.47i$ & $0.79-0.45i$ & $0.082-0.045i$ & $0.160-0.086i$ & $-0.115+0.063i$ \\
$\Db\Scs(\tfrac32^-)$ & $0.67-0.39i$ & $0.59-0.37i$ & $0.87-0.65i$ & $0.81-0.62i$ & $0.072-0.053i$ & $0.158-0.119i$ & $-0.129+0.096i$ \\
$\Dbs\Sc(\tfrac12^-)$ & $0.43-0.48i$ & $0.37-0.46i$ & $0.35-0.18i$ & $0.31-0.17i$ & $0.048-0.024i$ & $0.087-0.044i$ & $-0.035+0.017i$ \\
$\Dbs\Sc(\tfrac32^-)$ & $0.96-0.14i$ & $0.89-0.14i$ & $1.76-0.57i$ & $1.69-0.54i$ & $0.240-0.077i$ & $0.314-0.096i$ & $-0.176+0.057i$ \\
$\Dbs\Scs(\tfrac12^-)$ & $0.40-0.62i$ & $0.33-0.59i$ & $0.36-0.17i$ & $0.32-0.17i$ & $0.044-0.021i$ & $0.061-0.029i$ & $-0.040+0.019i$ \\
$\Dbs\Scs(\tfrac32^-)$ & $0.70-0.58i$ & $0.63-0.56i$ & $0.67-0.40i$ & $0.62-0.39i$ & $0.081-0.048i$ & $0.102-0.073i$ & $-0.075+0.045i$ \\
$\Dbs\Scs(\tfrac52^-)$ & $1.24$ & $1.17$ & $4.35$ & $4.22$ & $0.531$ & $0.775$ & $-0.489$ \\
\end{tabular}
\end{ruledtabular}
\end{table*}

\begin{figure*}[t]
\centering
\includegraphics[width=0.96\textwidth]{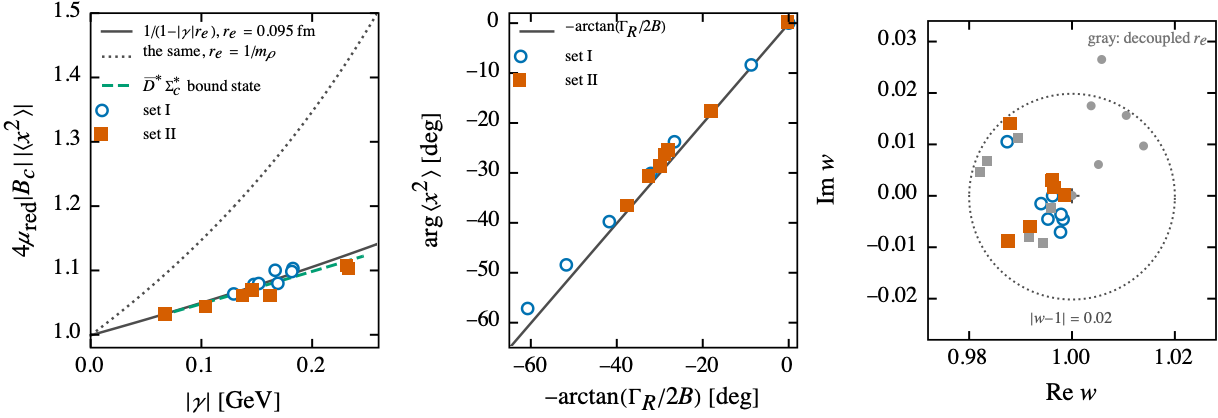}
\caption{Left: modulus of the ratio of $\MSD$ to the zero-range value $1/(4\mured B_c)$ for the seven states in
the two parameter sets, as a function of the modulus of the binding momentum. The solid line is the
effective-range factor of Eq.~\eqref{eq:ERE} for real $\gamma$ with $r_e=0.095\fm$, the dotted line the same
for $r_e=1/m_\rho$, and the dashed line the single-channel $\Dbs\Scs$ bound state with varied interaction
strength. Middle: phase of $\MSD$ compared with the phase of $1/B_c$. Right: the complex ratio
$w=4\mured B_c(1-\gamma r_e)\MSD$ with the complex effective range of the full amplitude (large symbols) and
with the real range of the decoupled channel (small gray symbols).}
\label{fig:zr}
\end{figure*}

The mean squared distance between the constituents of the dominant channel can be obtained from either the
baryon-number or the anticharm radius,
\begin{equation}
\MSD=\frac{\MSR_B}{\xi_B^2}=\frac{\MSR_{\bar c}}{\xi_M^2}.
\label{eq:MSD}
\end{equation}
The two determinations agree within $0.5\%$ for all states. The values of $\MSD$ and the radii are listed in
Table~\ref{tab:radii}. For a zero-range wave function, one obtains
\begin{equation}
\MSD_{\rm ZR}=\frac{1}{2\gamma^2}=\frac{1}{4\mured B_c}.
\label{eq:ZR}
\end{equation}
The computed values exceed Eq.~\eqref{eq:ZR} by $3$ to $11\%$ in modulus, whereas the phases agree within
$4^\circ$ (see Fig.~\ref{fig:zr}).

This excess can be understood as follows. The asymptotic normalization of the wave function of a shallow state is
determined by the residue at the pole, which depends on the effective range $r_e$ of the
amplitude~\cite{Weinberg:1965zz}. If the mean squared distance is dominated by the tail of the wave function,
one finds
\begin{equation}
\MSD\simeq\frac{1}{4\mured B_c}\,\frac{1}{1-\gamma r_e}.
\label{eq:ERE}
\end{equation}
The effective range is defined through the expansion $k\cot\delta_l=-1/a_0+\tfrac12r_ek^2$ of the amplitude
$f_l=-M_lT_{ll}/(4\pi\rs)=1/(k\cot\delta_l-ik)$ of the dominant channel. In the coupled-channel system, $r_e$ is
complex because the lower channels are open. For the full amplitude, we obtain $\Rl r_e=0.081$--$0.121\fm$ and
$|\Im r_e|\leq0.017\fm$ at the fourteen poles. When the other channels are switched off, the effective range
becomes real, $r_e=0.091$--$0.098\fm$. The expansion up to order $k^2$ reproduces the binding momenta of the poles
within $0.6\%$. To test Eq.~\eqref{eq:ERE} as a complex relation, we define the ratio
$w=4\mured B_c(1-\gamma r_e)\MSD$, which equals unity if Eq.~\eqref{eq:ERE} holds. With the effective range of the
full amplitude, we find $|w-1|\leq0.018$ for the fourteen cases of Table~\ref{tab:radii}, which cover the binding
momenta $|\gamma|=0.07$--$0.23\GeV$ (see Fig.~\ref{fig:zr}). With the real effective range of the decoupled
channel, $|w-1|\leq0.027$. At the level of $2\%$, the loop calculation with pointlike vertices is therefore
essentially determined by the binding energy, the width and an effective range of $0.1\fm$. The other channels
contribute mainly through the width in $B_c$ and through the complex effective range. Equation~\eqref{eq:ERE} is
the extension of the shallow-bound-state relation to a complex binding energy. As shown by the dashed line in
Fig.~\ref{fig:zr}, a single-channel bound state with varied interaction strength follows the same factor.

The effective range of the model is determined by the energy dependence of the kernel and by the regular part of
the loop function. It is unrelated to the range $1/m_\rho=0.25\fm$ of the vector-meson exchange, which is not
resolved by the present kernel. The values in Table~\ref{tab:radii} should therefore be understood as results
for the effective range of this model. The sensitivity to $r_e$ can be illustrated by two examples. For
$r_e=1/m_\rho$, the modulus of the factor in Eq.~\eqref{eq:ERE} is $1.22$, $1.40$ and $1.15$ for $P_c(4312)$,
$P_c(4440)$ and $P_c(4457)$ in set II, compared with the computed ratios $1.07$, $1.10$ and $1.04$. For the
deuteron, with $\gamma=0.2316\fm^{-1}$ and the triplet effective range $1.753\fm$, the factor is $1.68$, while
the experimental structure radius of $1.975\fm$~\cite{Garcon:2001} exceeds the zero-range value by a factor of $1.67$
in the mean squared radius. These examples, however, do not provide a bound. For $\gamma=u+iv$ and real $r_e$, the
factor exceeds unity in modulus only for $0<r_e<2u/|\gamma|^2$, whereas a negative or complex effective range can
reduce the modulus.

Since the radii of resonances are complex, their interpretation is not unique, and several interpretations have
been discussed in the literature~\cite{Berggren:1996,Sekihara:2013,YamagataSekihara:2011,Kamiya:2017}. Here we do
not adopt a particular interpretation and present both the real and the imaginary parts. In
Eq.~\eqref{eq:ZR}, the modulus and the phase have a simple meaning, $|\MSD|\simeq1/(4\mured|B_c|)$ and
$\arg\MSD\simeq-\arctan(\Gamma_R/2B)$. For $\Gamma_R\ll2B$, the imaginary part is small, and the real part can be
interpreted as a mean squared distance. For the states considered here, however, $\Gamma_R/2B=0.3$ to $1.8$,
except for the $\tfrac32^-$ $\Dbs\Sc$ state in set I. The modulus defines a characteristic length,
$|\MSD|^{1/2}=0.98$, $0.63$ and $1.36\fm$ for the three observed states in set II.

\subsection{Charge radii}
\label{subsec:constituents}

For pointlike constituents, the charge radius is given by the charge-weighted sum of the radii of the two
constituents,
\begin{equation}
\MSR_E\simeq\big[Q_M\xi_M^2+Q_B\xi_B^2\big]\MSD ,
\label{eq:rE}
\end{equation}
with $Q_M=-\tfrac23$ and $Q_B=\tfrac53$. The factor in brackets is $0.096$ for $\Db\Sc$ and $0.136$ for $\Dbs\Sc$.
Since the meson and the baryon carry charges of opposite sign and the heavier baryon is located closer to the
center of mass, the two terms largely cancel. The values in Table~\ref{tab:radii} satisfy Eq.~\eqref{eq:rE}
within $1\%$. The charm radius $\MSR_C$ is negative because the anticharm quark resides in the lighter
constituent. Two further contributions to the slope of $\GE$ are not included in these values.

The first contribution arises from the electric contact currents of Eq.~\eqref{eq:LE}. Their size can only be
estimated by dimensional analysis, $|\Ecal|\sim|V|/\Lambda^3$, which gives values between $15$ and
$90\GeV^{-5}$. Taking into account only the diagonal entry of the dominant channel, Eq.~\eqref{eq:drE} gives
$|\delta\MSR_E|=0.014\fm^2$ for $\Ecal=30\GeV^{-5}$ and $0.04\fm^2$ for $90\GeV^{-5}$ for each of the three
observed states in set II. These shifts correspond to $16$, $28$ and $5\%$ of the moduli of the values in
Table~\ref{tab:radii} for $30\GeV^{-5}$, and to $47$, $83$ and $16\%$ for $90\GeV^{-5}$. The effect is largest for
$P_c(4440)$, whose molecular charge radius is small owing to the cancellation in Eq.~\eqref{eq:rE}. The distance
$\MSD$, which is defined through the baryon-number and anticharm probes, is not affected by these terms.

The second contribution arises from the finite size of the constituents, which adds their charge-weighted radii,
including the radius of the neutral $\Db^{(*)0}$. We estimate this effect by vector-meson dominance at the quark
level, which respects the quark content of the constituents: the light quarks couple to the photon through the
$\rho$ and $\omega$ poles, and the charm quarks through the $\Jpsi$ pole. Since the charm contributions of the $c$
and the $\bar c$ cancel at $Q^2=0$ and the net charge of the light quarks of the $P_c^+$ is $+1$ in every channel,
the correction is the same for all states, $\delta\MSR_E=6/m_\rho^2=0.39\fm^2$. An explicit evaluation with these
form factors in all loops gives $\MSR_E=0.47-0.04i$, $0.43-0.02i$ and $0.63-0.08i\fm^2$ for the three observed
states in set II. At $Q^2=0.5\GeV^2$, the electric form factors are reduced to $0.48$, $0.51$ and $0.40$ from
$0.83$, $0.91$ and $0.69$ (see Fig.~\ref{fig:ff}). The same charge form factor is obtained if only the transverse
part of each vertex is dressed, so that the Ward identities remain satisfied. Since lattice QCD calculations find
charge radii of the charmed baryons smaller than that of the proton~\cite{Can:2014}, this estimate may overestimate
the effect. Nevertheless, it shows that the finite size of the constituents can exceed the molecular
contribution to the charge radius, $0.05$ to $0.24\fm^2$, and that it dominates the $Q^2$ dependence of $\GE$
above about $0.3\GeV^2$. The distance $\MSD$ between the centers of the constituents is not affected.

\section{Magnetic moments}
\label{sec:moments}

\subsection{Additive moments and loop results}
\label{subsec:additive}

\begin{table*}[t]
\caption{Magnetic moments in $\muN$. Second column: additive moment of the dominant channel. Third and fourth
columns: $\GM(0)$ in the two parameter sets with the constituent quark moments and without contact terms. All
other columns refer to the real part in set II. Fifth and sixth columns: light-quark part of the $\Sc^{(*)}$
moments scaled to the lattice results of Ref.~\cite{Can:2014} for the linear and the quadratic extrapolation.
Seventh and eighth columns: maximal shifts $\delta_{\rm ell}$ and $\delta_{\rm box}$ from the contact couplings, Eq.~\eqref{eq:box}. Ninth column: excess of the
loop normalization over the additive value. Last column: range of $\Rl\mu$ over these variations (see text).}
\label{tab:mu}
\begin{ruledtabular}
\begin{tabular}{lccccccccc}
 & & \multicolumn{2}{c}{$\GM(0)$} & \multicolumn{2}{c}{lattice $\mu_{\Sc}$} & \multicolumn{2}{c}{contact} & & \\
state & additive & set I & set II & linear & quadratic & $\delta_{\rm ell}$ & $\delta_{\rm box}$ & loops & range of $\Rl\mu$\\
\hline
$\Db\Sc(\tfrac12^-)$ & $+1.72$ & $+1.82+0.03i$ & $+1.74-0.01i$ & $+1.10$ & $+1.51$ & $0.16$ & $0.23$ & $0.09$ & $+0.8$ to $+2.0$ \\
$\Db\Scs(\tfrac32^-)$ & $+3.21$ & $+3.36+0.08i$ & $+3.24+0.03i$ & $+2.28$ & $+2.90$ & $0.21$ & $0.28$ & $0.15$ & $+1.8$ to $+3.5$ \\
$\Dbs\Sc(\tfrac12^-)$ & $-0.85$ & $-0.91-0.07i$ & $-0.95-0.05i$ & $-0.67$ & $-0.85$ & $0.25$ & $0.32$ & $0.07$ & $-1.3$ to $-0.2$ \\
$\Dbs\Sc(\tfrac32^-)$ & $+1.31$ & $+1.33-0.02i$ & $+1.26-0.03i$ & $+0.67$ & $+1.05$ & $0.27$ & $0.42$ & $0.07$ & $+0.2$ to $+1.8$ \\
$\Dbs\Scs(\tfrac12^-)$ & $+1.92$ & $+1.94+0.19i$ & $+2.14+0.17i$ & $+1.54$ & $+1.92$ & $0.21$ & $0.39$ & $0.16$ & $+0.9$ to $+2.5$ \\
$\Dbs\Scs(\tfrac32^-)$ & $+2.19$ & $+2.23+0.22i$ & $+2.41+0.17i$ & $+1.65$ & $+2.14$ & $0.31$ & $0.47$ & $0.13$ & $+1.0$ to $+2.9$ \\
$\Dbs\Scs(\tfrac52^-)$ & $+2.79$ & $+2.89$ & $+2.84$ & $+1.89$ & $+2.50$ & $0.24$ & $0.41$ & $0.08$ & $+1.4$ to $+3.4$ \\
\end{tabular}
\end{ruledtabular}
\end{table*}

The magnetic moment $\mu=\GM(0)\,\muN$ is complex at a resonance pole, and our statements on signs and
ranges refer to its real part. For an $S$-wave meson--baryon state, the additive moment $\mu_{\rm add}$ is the
diagonal element of $\mathcal M^{(1)}$ in Eq.~\eqref{eq:WE}. With the isospin-averaged moments of
Table~\ref{tab:moments}, it equals $\mu_{\Sc}$ for the $\Db\Sc$ state,
$\tfrac23\mu_{\Dbs}-\tfrac13\mu_{\Sc}=-0.85\,\muN$ for the $\tfrac12^-$ $\Dbs\Sc$ state and
$\mu_{\Dbs}+\mu_{\Sc}=1.31\,\muN$ for the $\tfrac32^-$ state (second column of Table~\ref{tab:mu}), as in the
quark-model calculations of Refs.~\cite{Wang:2016dzu,Li:2021ryu}.

The full results in the third and fourth columns of Table~\ref{tab:mu} differ from the additive values by
$1$ to $6\%$ in set I and by $1$ to $11\%$ in set II. Two effects contribute to the results. First, the ratio of the diagonal
magnetic and electric loops of the dominant channel, which equals unity in the nonrelativistic limit, ranges from
$1.02$ to $1.07$ at the fourteen poles in dimensional regularization. With a three-momentum cutoff that
reproduces the subtraction constant at threshold, it is $1.006$ to $1.024$ for the observed states, so that this
effect depends on loop momenta of the order of the hadron masses. Second, the transition moments and the
subdominant channels contribute $-0.06$ to $0$ in set I and $-0.07$ to $+0.17\,\muN$ in set II, where $r\neq1$
couples the $\Db^{(*)}\Sc^{(*)}$ channels directly. The imaginary parts of $\GM(0)$ are $-0.01$, $-0.05$ and
$-0.03$ for the observed states in set II and below $0.22$ in modulus otherwise. The additive moments therefore
dominate also in the present approach.

\subsection{Sensitivity on the inputs}
\label{subsec:budget}

The variations of the inputs exceed these loop corrections. We consider four sources as follows.

\emph{Constituent moments.} In the $I=\tfrac12$ average, the lattice results for the
$\Sc$~\cite{Can:2014} (Table~\ref{tab:moments}) lie $36\%$ below the quark-model value for the linear
extrapolation in $m_\pi^2$ and $13\%$ below for the quadratic one. Scaling the light-quark parts of all
$\Sc^{(*)}$ moments and transition moments by $0.67$ and $0.88$, with the $\Sc$ and the $\Scs$ related by HQSS,
gives the fifth and sixth columns of Table~\ref{tab:mu}. The meson transition moments extracted from the
radiative widths of $D^{*+}\to D^+\gamma$ and $\Jpsi\to\etac\gamma$ are $14$ and $20\%$ below the constituent-quark
values (Appendix~\ref{app:moments}). A $30\%$ reduction of the meson-line contribution changes $\mu$ by at most
$0.13\,\muN$.

\emph{Contact couplings and loop normalization.} The maximal shifts $\delta_{\rm ell}$ and $\delta_{\rm box}$
for the bounds of Eq.~\eqref{eq:domain} are listed in Table~\ref{tab:mu} for set II. In set I they are $0.10$ to
$0.16\,\muN$ and $0.15$ to $0.30\,\muN$. Both scale linearly with the bounds, and the exchange-current estimate of
Appendix~\ref{app:fivepoint} gives at most $0.07\,\muN$. The excess of the diagonal loops over the additive
value, $0.07$ to $0.16\,\muN$ in set II, is included as a one-sided variation. It corresponds to a shift of a
diagonal contact coupling by $8$ to $17\GeV^{-4}$ for the observed states, within the bounds.

\emph{Regularization.} In a refit, the common subtraction constant $a$ was shifted by $\pm0.2$ and the
kernel was multiplied by a factor $\lambda_V$, with $\lambda_V$, $\delta$ and $r$ determined from the masses and
$\kappa$ from the widths. This gives $\lambda_V=0.57$ for $a=-2.43$ and $\lambda_V=3.27$ for $a=-2.03$
($\chi^2=1.6$ and $1.9$). Without contact terms, the moments of the observed states change by at most
$0.024\,\muN$ and $|4\mured B_c\MSD|$ by at most $0.016$. At fixed pole positions, these quantities are thus
insensitive to the subtraction constant, although the kernel strength changes by a factor of $5.7$. The contact
terms behave differently, because the amplitudes $\psi_l$ of the dominant channels change by factors of $1.4$ to
$1.6$ and $0.4$ to $0.6$, respectively. For the fixed bounds of Eq.~\eqref{eq:domain}, $\delta_{\rm ell}$ of the
observed states changes from $0.16$--$0.27\,\muN$ to $0.36$--$0.59\,\muN$ and $0.05$--$0.10\,\muN$. For bounds
scaled with the kernel strength, as required by the estimate $|\bar c|\sim|V|/\Lambda^2$, it is
$0.20$--$0.34\,\muN$ and $0.18$--$0.31\,\muN$. The shifts $\delta_{\rm ell}$ and $\delta_{\rm box}$ thus represent the dimensional estimate within a
given regularization scheme, consistent with Eq.~\eqref{eq:matching}.

Since the constituent moments and the loop normalization shift the moments in one direction only, we do not
add the variations in quadrature. The last column of Table~\ref{tab:mu} gives the range of $\Rl\mu$, from the
smallest modulus obtained with the input variations, reduced by $\delta_{\rm box}$ and by the loop excess, to the
largest modulus, increased by $\delta_{\rm box}$. With $\delta_{\rm ell}$, the ranges of the observed states
become $+0.8$ to $+1.9$, $-1.2$ to $-0.3$ and $+0.3$ to $+1.7\,\muN$. These ranges represent model variations
rather than statistical confidence intervals. All real parts retain their signs. For the two $\Dbs\Sc$ states,
however, the ranges reach zero if the bounds are increased by factors of $1.7$ for $P_c(4440)$ and $1.4$ for
$P_c(4457)$. The signs are therefore determined only for contact couplings below about $40\GeV^{-4}$ in the
present scheme, a value within the range of the dimensional estimate.

\subsection{Spin assignment and comparison with other calculations}
\label{subsec:compare}

\begin{table*}[t]
\caption{Real parts of the magnetic moments of the three observed states in $\muN$ for the two spin orderings
of the $\Dbs\Sc$ states, compared with other calculations for molecular configurations. This work: central
value and range as in Table~\ref{tab:mu}. Reference~\cite{Wang:2016dzu}: quark model.
Reference~\cite{Li:2021ryu}: molecular picture with a single channel, with coupled channels, and with $D$-wave
components. Reference~\cite{LiHS:2024a}: chiral perturbation theory. Reference~\cite{Ozdem:2021}: light-cone QCD
sum rules with a molecular current.}
\label{tab:compare}
\begin{ruledtabular}
\begin{tabular}{llcccccc}
state & assignment & \multicolumn{2}{c}{this work} & Ref.~\cite{Wang:2016dzu} & Ref.~\cite{Li:2021ryu} & Ref.~\cite{LiHS:2024a} & Ref.~\cite{Ozdem:2021}\\
 & & central & range & & & & \\
\hline
$P_c(4312)$ & $\Db\Sc$, $\tfrac12^-$ & $+1.74$ & $+0.8$ to $+2.0$ & $1.76$ & $1.74$, $1.63$, $1.62$ & $1.53$ & \\
\multicolumn{8}{l}{Ordering A}\\
$P_c(4440)$ & $\Dbs\Sc$, $\tfrac12^-$ & $-0.95$ & $-1.3$ to $-0.2$ & $-0.86$ & $-0.83$, $-0.97$, $-0.98$ & & \\
$P_c(4457)$ & $\Dbs\Sc$, $\tfrac32^-$ & $+1.26$ & $+0.2$ to $+1.8$ & $1.36$ & $1.37$, $1.12$, $1.15$ & & \\
\multicolumn{8}{l}{Ordering B}\\
$P_c(4440)$ & $\Dbs\Sc$, $\tfrac32^-$ & $+1.35$ & $+0.2$ to $+1.8$ & $1.36$ & & & $3.49^{+1.49}_{-1.31}$\\
$P_c(4457)$ & $\Dbs\Sc$, $\tfrac12^-$ & $-0.89$ & $-1.0$ to $-0.4$ & $-0.86$ & & & $2.78^{+0.94}_{-0.83}$\\
\end{tabular}
\end{ruledtabular}
\end{table*}

The additive moment of a $\Dbs\Sc$ state is negative for $J=\tfrac12$ if $\langle\mu_{\Dbs}\rangle<0$
and $\mu_{\Sc}>0$, and positive for $J=\tfrac32$ if, in addition, $\mu_{\Sc}>|\langle\mu_{\Dbs}\rangle|$. Both
conditions hold for all inputs considered here, with $\mu_{\Sc}$ between $1.10$ and $1.72\,\muN$ and
$|\langle\mu_{\Dbs}\rangle|$ between $0.29$ and $0.42\,\muN$, so that the signs follow from the constituent
moments and the one-body current. For contact couplings within the bounds of Eq.~\eqref{eq:domain}, the two spin
assignments give real parts of opposite sign, and the sign of the moment would determine the spin of a $\Dbs\Sc$
molecule.

Table~\ref{tab:compare} compares the two orderings. Since the kernel of Sec.~\ref{subsec:model} cannot be
fitted to the data for ordering B, we place the two poles at the observed masses by shifting the subtraction
constant of the $\Dbs\Sc$ channel in set I by $-0.092$ for $J=\tfrac32$ and by $+0.046$ for $J=\tfrac12$. This
breaks HQSS and serves only to evaluate the loops at the required pole positions. The resulting moments,
$+1.35-0.01i$ and $-0.89-0.07i\,\muN$, differ in their real parts from those of ordering A for the same spin by
less than $0.1\,\muN$, so that the moments of the two states are interchanged. The ranges for ordering B were
computed at its own pole positions.

The quark-model values of Ref.~\cite{Wang:2016dzu} for molecular configurations are additive moments and
agree with the second column of Table~\ref{tab:mu} within $0.05\,\muN$. The closest previous calculation of the
static moments, Ref.~\cite{Li:2021ryu}, treats the states as coupled-channel molecules with $D$-wave components and
constituent quark moments. Its values differ from our central values by at most $0.14\,\muN$, and the $D$-wave
components change the moments by less than $0.03\,\muN$. Chiral perturbation theory gives $1.53\,\muN$ for the
$\Db\Sc$ state~\cite{LiHS:2024a}. All these values lie within our ranges. In contrast, the light-cone sum rules of
Ref.~\cite{Ozdem:2021} with a molecular interpolating current and ordering B yield larger positive moments for
both states, for the $\tfrac12^-$ state opposite in sign to the additive value. Compact configurations give still
different moments~\cite{Wang:2016dzu,OrtizPacheco:2019,Ozdem:2021,Ozdem:2026cmp}. Within the molecular picture,
the sign of the moment of either $\Dbs\Sc$ state would therefore constrain the spin assignment and discriminate
among the calculations listed in Table~\ref{tab:compare}.

\section{Sensitivity on the observed masses and widths}
\label{sec:measured}

\begin{table*}[t]
\caption{Mean squared distance and charge radius of the three observed states in fm$^2$ at the poles of set II
and their ranges. $B$ is the binding energy in MeV at the central mass. ``Mass'': the mass is varied within its
experimental uncertainty at the width of set II. ``Mass and width'': the mass and the width are varied within the
experimental uncertainties. For $P_c(4457)$ the width of set II lies below the experimental interval, so that the two
ranges are not nested.}
\label{tab:observed}
\begin{ruledtabular}
\begin{tabular}{lcccccccc}
 & & & \multicolumn{2}{c}{mass} & \multicolumn{2}{c}{mass and width} & & mass\\
state & $B$ & $\MSD$ & $\Rl\MSD$ & $\Im\MSD$ & $\Rl\MSD$ & $\Im\MSD$ & $\MSR_E$ & $\Rl\MSR_E$\\
\hline
$\Db\Sc(\tfrac12^-)$ & $8.8$ & $0.85-0.47i$ & $0.69$ to $0.99$ & $-1.7$ to $-0.40$ & $0.37$ to $2.2$ & $-2.4$ to $-0.21$ & $0.082-0.045i$ & $0.067$ to $0.096$ \\
$\Dbs\Sc(\tfrac12^-)$ & $21.7$ & $0.35-0.18i$ & $0.31$ to $0.39$ & $-0.24$ to $-0.12$ & $0.28$ to $0.52$ & $-0.26$ to $-0.049$ & $0.048-0.024i$ & $0.042$ to $0.053$ \\
$\Dbs\Sc(\tfrac32^-)$ & $4.7$ & $1.76-0.57i$ & $1.3$ to $3.0$ & $-5.3$ to $-0.31$ & $0.14$ to $2.5$ & $-4.6$ to $-0.36$ & $0.240-0.077i$ & $0.18$ to $0.41$ \\
\end{tabular}
\end{ruledtabular}
\end{table*}

The observed masses have uncertainties of $^{+6.8}_{-0.9}$, $^{+4.3}_{-4.9}$ and $^{+4.1}_{-1.8}\MeV$,
which are not small compared with the binding energies of $8.8$, $21.7$ and $4.7\MeV$. Since $\MSD\propto1/B_c$,
they dominate the variation of the radii. When each pole is shifted to the ends of the observed mass range by
varying the subtraction constant of the dominant channel, the ratio of the computed $\MSD$ to $1/(4\mured B_c)$
remains between $1.02$ and $1.16$ in modulus for binding energies between $0.6$ and $26.6\MeV$, in agreement with
Eq.~\eqref{eq:ERE}. The ranges in Table~\ref{tab:observed} are therefore obtained from the zero-range relation
multiplied by the computed ratio at the central pole. With the widths of set II, $|\MSD|$ changes by $-7$ to
$+88\%$, $-15$ to $+17\%$ and $-26$ to $+205\%$ for the three states. At fixed width, the real part of
Eq.~\eqref{eq:ZR} is maximal, $1/(4\mured\Gamma_R)$, at $B=\Gamma_R/2$ and vanishes at threshold, where the
imaginary part approaches $-1/(2\mured\Gamma_R)$. If the width is also varied within its uncertainty, the real
parts cover $0.4$--$2.2$, $0.3$--$0.5$ and $0.1$--$2.5\fm^2$. Only for $P_c(4440)$ is $\MSD$ thus determined
within a factor of two. The magnetic moments change by at most $0.06\,\muN$ over the mass ranges.

The uncertainty of $\kappa$ has a smaller effect. At the ends of its interval, with the other parameters
redetermined from the masses, the widths of the observed states are $6.6$, $16.7$ and $2.4\MeV$ and $13.4$,
$30.2$ and $3.6\MeV$. The moduli of $\MSD$ then change by $8$, $6$ and $2\%$, and the real parts of the magnetic
moments by less than $0.01\,\muN$. Since the spin does not enter Eq.~\eqref{eq:ERE}, the ranges in
Table~\ref{tab:observed} apply to both spin orderings of the $\Dbs\Sc$ states.

The calculation employs isospin-averaged masses, whereas the thresholds of the two charge states differ by
$6.1\MeV$ for $\Db\Sc$ and $4.7\MeV$ for $\Dbs\Sc$, which is comparable with the binding energies of $P_c(4312)$
and $P_c(4457)$. Isospin breaking in $P_c(4457)$ has been proposed as a test of its molecular
nature~\cite{Guo:2019iso}. Instead of a calculation in the charge basis, we estimate the effect in the zero-range
limit, with the pole held fixed and the tails of the two charge components determined by their physical
thresholds. The real part of $\MSD$ then changes by $-6$, $-2$ and $+7\%$ for the three observed states. The
charge radius is more sensitive, because the neutral-meson component has the lower threshold and hence the longer
tail, and its charges do not cancel. The real part of $\MSR_E$ increases by $8$, $2$ and $30\%$.

Finally, the constituents are treated as stable, whereas the isospin-weighted widths are
$\Gamma_{\Sc}=2.0\MeV$ and $\Gamma_{\Scs}=15.6\MeV$~\cite{ParticleDataGroup:2026}. A constant constituent width
$\Gamma_{\rm con}$ shifts the threshold and, to leading order, the pole by $-i\Gamma_{\rm con}/2$, so that $B_c$
and Eq.~\eqref{eq:ERE} remain unchanged and the width of the state becomes $\Gamma_R+\Gamma_{\rm con}$.
Corrections from the energy dependence of the self-energy and from the three-body channels $\Db^{(*)}\Lc\pi$ are not
included. Their size is indicated by $\Gamma_{\rm con}/(2|B_c|)$, which is $0.10$, $0.04$ and $0.20$ for the
observed states and $0.3$ to $3.8$ for the four states with a $\Scs$. For the latter, the pole positions and radii
should therefore be regarded as reference values for stable constituents.

\section{Open channels and the triangle singularity}
\label{sec:open}

\begin{figure*}[t]
\centering
\includegraphics[width=0.9\textwidth]{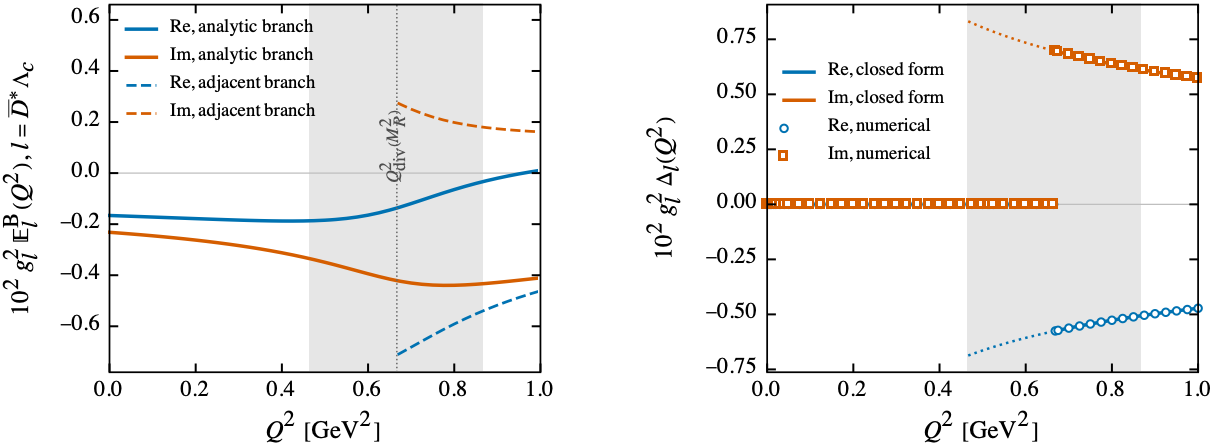}
\caption{Left: contribution $g_l^2\,\EB_l$ of the $\Dbs\Lc$ channel to the baryon-number form factor of
$P_c(4312)$ in set II. Solid lines: the branch that is analytic in $Q^2$ and contains $Q^2=0$. Dashed lines: the
adjacent branch, shown above the triangle point $Q^2_{\rm div}(M_R^2)$. Right: difference of the two branches, 
$g_l^2\Delta_l$. Symbols are the difference of the two numerical continuations, lines the closed form
\eqref{eq:band}. The shaded area is the triangle interval \eqref{eq:trint}.}
\label{fig:open}
\end{figure*}

The photon loops of a channel $l$ that is open at the energy $\rs$ are singular at the spacelike momentum
transfer~\cite{Sekihara:2010uz}
\begin{equation}
Q^2_{\rm div}(s)=\frac{\lambda(s,\mst^2,\msp^2)}{\msp^2}=\frac{4q_l^2(s)\,s}{\msp^2},
\label{eq:Qdiv}
\end{equation}
where $\mst$ is the mass of the constituent struck by the photon, $\msp$ that of the spectator, and $\lambda$ is
the K\"all\'en function. This singularity is the leading Landau singularity of the triangle
graph~\cite{Landau:1959fi,Coleman:1965xm,Guo:2019twa}. It corresponds to a process in which the state decays into
the channel, the photon reverses the momentum of the struck constituent in the rest frame of the spectator, and
the two constituents recombine with all three internal lines on shell. For a closed channel, the same expression
gives the anomalous threshold of a bound-state form factor at timelike momentum transfer~\cite{Karplus:1958zz}.

At the pole, $s=s'=\zR^2$, the triangle point lies at a complex momentum transfer below the real axis, for
example at $Q^2_{\rm div}(\zR^2)=(0.67-0.20i)\GeV^2$ for the baryon loop of the $\Dbs\Lc$ channel of $P_c(4312)$
in set II. Consequently, the double-pole residue \eqref{eq:residue} is an analytic function of $Q^2$ in a
neighborhood of the real axis. We define the form factor $\Fan$ as the branch of the residue that contains
$Q^2=0$. Following Ref.~\cite{Sekihara:2010uz}, it is obtained by the continuation in the energy at $Q^2\to0$,
followed by the continuation in $Q^2$ along the real axis at fixed $\zR$. This branch satisfies the Ward
identities and is normalized to the charge, and its Taylor coefficients at $Q^2=0$ give the radii and the
moments. The triangle point is a logarithmic branch point of the residue. The continuation of $\Fan$ around this
point leads to the adjacent branch, on which the diagonal electric loops $\Ean_l$ are replaced by
$\Ean_l+\Delta_l$ with
\begin{equation}
\Delta_l(Q^2)=\frac{M_l\,\big[2(\zR^2+\mst^2-\msp^2)+Q^2\big]}{2Q\,(4\zR^2+Q^2)}
\label{eq:band}
\end{equation}
(see Appendix~\ref{app:loops}). For the magnetic loops with the photon on the meson and the baryon line, the
corresponding differences are $M_l\sqrt{m_lm_{l'}}/(2QE)$ and $M_lM_{l'}/(2QE)$. These analytic expressions agree
with the difference of the two numerical continuations to machine precision (see Fig.~\ref{fig:open}). Although
the adjacent branch also satisfies the Ward identities, it is not normalized, since $\Delta_l$ grows as $1/Q$ for
$Q^2\to0$. The requirement of regularity at $Q^2=0$ therefore selects $\Fan$.

The relation between the two branches and the amplitude at real energies is as follows. For a given real
$Q^2>0$, each loop of an open channel has a logarithmic branch point at the real energy $E_{\rm tr}(Q^2)$ defined
by $Q^2_{\rm div}(E_{\rm tr}^2)=Q^2$. The pole can be reached from real energies on either side of this point. The
continuation in the energy at fixed $Q^2$ yields $\Ean_l$ if it starts at $\rs>E_{\rm tr}$, and $\Ean_l+\Delta_l$
if it starts at $\rs<E_{\rm tr}$. For $Q^2<Q^2_{\rm div}(M_R^2)$, the point $E_{\rm tr}$ lies below $M_R$,
whereas for larger $Q^2$ it lies above $M_R$. The value of the residue that is connected to the amplitude at the
resonance energy by a continuation at fixed $Q^2$ is therefore
\begin{equation}
\Fcon(Q^2)=\Fan(Q^2)+\sum\theta\big(Q^2-Q^2_{\rm div}(M_R^2)\big)\,\Delta F(Q^2),
\label{eq:Fcon}
\end{equation}
where the sum runs over the loops of the open channels, each with its own triangle point and with the difference
$\Delta F$ that follows from Eq.~\eqref{eq:band} and the couplings and charges of Eq.~\eqref{eq:final}. Below the
lowest triangle point, $\Fcon=\Fan$, so that the charge, the radii and the moments coincide. Since $\Fcon$ is a
piecewise selection of branches and not an analytic function of $Q^2$, we do not use it to define the form
factor. Instead, it specifies which branch of the residue multiplies the double pole of the amplitude at real
energies close to $M_R$. If the triangle point lies within the half width of the state,
\begin{equation}
Q^2_{\rm div}\big((M_R-\tfrac12\Gamma_R)^2\big)<Q^2<Q^2_{\rm div}\big((M_R+\tfrac12\Gamma_R)^2\big),
\label{eq:trint}
\end{equation}
the branch point $E_{\rm tr}(Q^2)$ lies inside the resonance peak, and the amplitude on either side of it is
connected to a different branch.

Among the three observed states, only $P_c(4312)$ has a triangle point below $1\GeV^2$. In set II, its $\Dbs\Lc$
channel is open by $16.9\MeV$, with $X_l=-0.002-0.002i$. The triangle points are located at
$Q^2_{\rm div}(M_R^2)=0.51\GeV^2$ for the meson loop and $0.67\GeV^2$ for the baryon loop, and the corresponding
intervals \eqref{eq:trint} are $0.36<Q^2<0.67\GeV^2$ and $0.47<Q^2<0.87\GeV^2$. The two branches differ by at most
$0.009$ in $\GE$, which corresponds to $1.2\%$, and by $0.009$ in $\GM$. In set I, the channel is open by
$11.6\MeV$ with $|X_l|=0.010$, and the difference reaches at most $0.035$. For the other states, the lowest
triangle points of the diagonal loops are located at $1.3$, $1.6$, $2.0$, $2.6$ and $3.4\GeV^2$ in set II. The
lowest triangle point of a magnetic transition loop between two open channels lies at $2.1\GeV^2$. The difference
between the two branches is thus smaller than the variation of the form factors with the parameter set, the
structure of the constituents and the contact terms.

The on-shell formalism of Ref.~\cite{Briceno:2020vgp} defines the form factor of a resonance by the same
double-pole residue. The continuation of the triangle function to the second sheets given in Eq.~(24) of
Ref.~\cite{Briceno:2020vgp} agrees with ours at real energies (see Appendix~\ref{app:loops}). In that work,
however, the form factor is obtained by a continuation in the energies at fixed real $Q^2$. When started at
$\rs=M_R$, this continuation yields $\Fcon$, which coincides with $\Fan$ below the triangle point but requires the
specification given here above it. The two branches exist for the form factor of any resonance with an open decay
channel.

In the $\Lambda(1405)$ study of Ref.~\cite{Sekihara:2010uz}, the $\pi\Sigma$ channel is open, and for the
higher pole Eq.~\eqref{eq:Qdiv} gives $0.17\GeV^2$ for the loop with the photon on the pion. This singularity
produces the structure near $0.2\GeV^2$ in the $\pi^+\Sigma^-$ component of the electric form factor shown in that
reference. Since the $\pi$ and the $\Sigma$ have zero hypercharge, their charges are pure isovector charges. For
an isoscalar state with isospin-averaged masses, the $\pi^+\Sigma^-$ and $\pi^-\Sigma^+$ components therefore
cancel at every $Q^2$ and on both branches. The baryon-number, strangeness and magnetic currents couple only to
the $\Sigma$. With the light pion as spectator, the corresponding triangle point lies at $12.7\GeV^2$, far outside
the range of $Q^2$ considered there. In $P_c(4312)$, the $\Dbs\Lc$ channel has the single charge state
$\bar D^{*0}\Lambda_c^+$, so that no such cancellation occurs.

Finally, we note that this singularity is of
the same type as the triangle singularities proposed as the origin of the $P_c$
signals~\cite{Guo:2015umn,Liu:2015fea,Mikhasenko:2015vca,Guo:2016bkl,Bayar:2016ftu}, but it appears in a different
graph and in a different kinematic variable. Since it presupposes the existence of the pole, it does not bear on
the question of whether the observed peaks are of kinematical origin.

\section{Summary}
\label{sec:summary}

In this work, we have studied the electromagnetic form factors of the hidden-charm pentaquarks as
$\Db^{(*)}\Sc^{(*)}$ molecules in a coupled-channel unitary approach with a vector-meson exchange kernel that
respects HQSS. The form factors are obtained from the residues at the double pole of the photon-coupled
amplitude. We have constructed the photon--four-hadron vertex of the HQSS channel space. For a charged state, the
vertex of the gauged energy-dependent kernel is required, and the charge normalization then coincides with the
compositeness sum rule. The one-body and contact magnetic operators connect the channels related by HQSS. Our
results are obtained for stable, pointlike constituents and isospin-averaged masses.

We have found that the complex mean squared distance between the constituents satisfies
$\MSD=[4\mured B_c(1-\gamma r_e)]^{-1}$ within $2\%$, where $B_c$ is the complex binding energy and $r_e$ the
complex effective range of the dominant channel, with $\Rl r_e=0.08$--$0.12\fm$. With the parameters fitted to
the three observed masses and widths, we obtain $\MSD=0.85-0.47i$, $0.35-0.18i$ and $1.76-0.57i\fm^2$ for
$P_c(4312)$, $P_c(4440)$ and $P_c(4457)$, respectively. The uncertainties of these masses change the
modulus of $\MSD$ by $-7$ to $+88\%$, $-15$ to $+17\%$ and $-26$ to $+205\%$. The molecular contribution to the
charge radius is small, $0.05$ to $0.24\fm^2$ in the real part, because the meson and the baryon carry charges of
opposite sign. Electric contact currents of the size suggested by dimensional analysis change this contribution
by $0.01$ to $0.04\fm^2$, while the finite size of the constituents adds $0.39\fm^2$ in a vector-meson dominance
estimate.

The real parts of the magnetic moments agree with the additive moments of the constituents within $1$--$11\%$.
Using constituent quark moments, we obtain $\Rl\mu=1.74$, $-0.95$ and $1.26\,\muN$, with imaginary parts below
$0.06\,\muN$. Taking into account the lattice results for the $\Sc$ moments, contact couplings within the bounds
from dimensional analysis and the loop normalization, we obtain the ranges $0.8$ to $2.0$, $-1.3$ to $-0.2$ and
$0.2$ to $1.8\,\muN$. These results correspond to the spin assignment $\tfrac12^-$ for $P_c(4440)$ and
$\tfrac32^-$ for $P_c(4457)$. For the opposite ordering, the moments of the two states are interchanged. The
opposite signs of the moments of the two $\Dbs\Sc$ states persist for the constituent moments considered here and
for contact couplings up to $1.4$ times the adopted bounds.

We have also shown that the photon loops of an open channel have a triangle singularity at spacelike momentum
transfer. At the pole, this singularity becomes a branch point at complex $Q^2$, and the form factor is defined
as the branch of the residue that is analytic at $Q^2=0$. Beyond the triangle point associated with the resonance
energy, the value connected to the amplitude at real energies lies on the adjacent branch, which differs by a
term given in closed form. The charge, the radii and the moments are the same on both branches. For the observed
states and $Q^2\leq1\GeV^2$, this effect concerns only $P_c(4312)$, for which the two branches differ by $1\%$ in
set II and $4\%$ in set I.

The quantities at the poles are complex and are not direct observables, and the present model contains only the
molecular component. The largest variation of the magnetic moments arises from the moments of the $\Sc^{(*)}$ and
$\Dbs$, so that lattice QCD results at the physical pion mass would reduce the present ranges. A determination of
the effective range of the $\Db^{(*)}\Sc$ amplitudes from line shapes would test the value of $0.1\fm$ obtained
with the present kernel. The inclusion of one-pion exchange with $D$ waves, a calculation in the charge basis and
the effects of the widths of the constituents are left for future work. The partner states, including a
$\tfrac52^-$ $\Dbs\Scs$ state $2\MeV$ below its threshold in set II, are predictions of the model for stable
constituents.

\section*{Data availability}
There are no publicly available research data or software supporting this manuscript. Requests for further
information or data should be sent to the authors.

\begin{acknowledgments}
This research project is supported by National Research Council of Thailand (NRCT): (Contract No.\ N41A680287).
DS is supported by Thailand NSRF via PMU-B [grant number B39G680009]. DS has also received funding support from
the Fundamental Fund of Khon Kaen University. Generative AI tools (Claude, Anthropic, and ChatGPT, OpenAI) were
used to assist with the drafting and grammatical correction of the manuscript text and with numerical
implementation. The authors directed this assistance, verified all scientific content and equations against
their own calculations, and take full responsibility for the paper.
\end{acknowledgments}

\appendix

\section{Photon loops and their continuation}
\label{app:loops}

The triangle loops $D^\mu_M$ and $D^\mu_B$ in Eq.~\eqref{eq:Xdef} are evaluated with Feynman parameters, as in
Appendix~B of Ref.~\cite{Sekihara:2010uz}. With $\iint\equiv\int_0^1\dd x\int_0^{1-x}\dd y$ and $z=1-x-y$, the
loop functions in Eq.~\eqref{eq:final} at $\rs=\rsp=\zR$ are given by
\begin{equation}
\begin{aligned}
\EM_l&=\frac{2M_lE}{16\pi^2}\iint\frac{2z}{\mathcal D^{\rm M}_{ll}},&
\MM_{l'l}&=\frac{2M_l\,2\sqrt{m_lm_{l'}}}{16\pi^2}\iint\frac{1}{\mathcal D^{\rm M}_{l'l}},\\
\EB_l&=\frac{2M_lE}{16\pi^2}\iint\frac{2x}{\mathcal D^{\rm B}_{ll}},&
\MB_{l'l}&=\frac{2M_l\,2M_{l'}}{16\pi^2}\iint\frac{1}{\mathcal D^{\rm B}_{l'l}},
\end{aligned}
\label{eq:loops}
\end{equation}
with the denominators
\begin{equation}
\begin{aligned}
\mathcal D^{\rm M}_{l'l}&=xm_l^2+ym_{l'}^2+zM_l^2+xyQ^2-z(1-z)\zR^2,\\
\mathcal D^{\rm B}_{l'l}&=xm_l^2+yM_l^2+zM_{l'}^2+yzQ^2-x(1-x)\zR^2.
\end{aligned}
\label{eq:denoms}
\end{equation}
The integrals over the real simplex define the loop functions on the first sheet in $s$ and $s'$. For an open
channel, the second sheet is reached by adding the discontinuity across the unitarity cut, which follows from the
Cutkosky rule~\cite{Cutkosky:1960,Veltman:1963th}. If both channels of a loop are open,
\begin{equation}
F^{\rm II,II}=F^{\rm I,I}+{\rm disc}_sF\big|_{+i\varepsilon}+{\rm disc}_{s'}F\big|_{-i\varepsilon},
\label{eq:sheetIIII}
\end{equation}
where the uncut propagator carries the Feynman prescription in the first discontinuity and the conjugate
prescription in the second. For real energies above both thresholds, Eq.~\eqref{eq:sheetIIII} reduces to
$F^{\rm II,II}=F-2i\,\Im F$, which is the continuation of the triangle function given in Eq.~(24) of
Ref.~\cite{Briceno:2020vgp}.

In the discontinuity in $s$, the two cut lines are on shell, and the uncut propagator is linear in the cosine of
the angle between the on-shell constituent and the photon in the rest frame of $P$, $\alpha\mp\beta\cos\theta$.
The angular integration yields the logarithm $L=\ln w$ with $w=(\alpha+\beta)/(\alpha-\beta)$. For the scalar
triangle $S=\iint1/\mathcal D$, where $\mathcal D$ is one of the denominators in Eq.~\eqref{eq:denoms}, we obtain
\begin{equation}
{\rm disc}_sS=-\frac{iq_l}{\rs}\,\frac{2\pi}{\beta}\,L ,
\label{eq:discs}
\end{equation}
where $q_l$ denotes the momentum on the sheet of the pole, with $\Rl q_l>0$ for an open channel. For a diagonal
loop at $s=s'$, one has $\beta=2q_lQE/\rs$, $\alpha=-Q^2\omega_{\rm sp}/\rs$ and
\begin{equation}
w=-\frac{1-z^\star}{1+z^\star},\qquad z^\star=\frac{Q\,\omega_{\rm sp}}{2q_lE},
\label{eq:zstar}
\end{equation}
where $\omega_{\rm sp}=(s+\msp^2-\mst^2)/(2\rs)$ is the energy of the spectator. The zero of $w$ at $z^\star=1$
corresponds to the triangle point \eqref{eq:Qdiv}. Since $\Im z^\star>0$ at the pole, $w$ does not cross the cut
of the principal value ${\rm Ln}$ for real $Q^2>0$. For the continuation in the energy at $Q^2\to0$ followed by the
continuation in $Q^2$, which defines $\Fan$, the logarithms of the terms cut in $s$ and in $s'$ are
\begin{equation}
L_s={\rm Ln}\,w-2\pi i,\qquad L_{s'}={\rm Ln}\,w
\label{eq:Lan}
\end{equation}
for all $Q^2$. For the continuation in the energy at fixed $Q^2$ from $\rs=M_R$, they coincide with
Eq.~\eqref{eq:Lan} below the triangle point, whereas $L_s=L_{s'}={\rm Ln}\,w$ above it. The two branches therefore
differ by $\delta L=2\pi i$ in the term cut in $s$. In Eq.~\eqref{eq:discs}, the momentum $q_l$ then cancels, and
$\delta S=2\pi^2/(QE)$. The same steps applied to the charge vertices give Eq.~\eqref{eq:band} for the electric
loops, while the prefactors in Eq.~\eqref{eq:loops} give the differences $M_l\sqrt{m_lm_{l'}}/(2QE)$ and
$M_lM_{l'}/(2QE)$ for the magnetic loops. For a transition loop between two open channels, in which the mass of
the struck constituent changes from $\mst$ to $\mst'$, the triangle point is located at
$Q^2_{\rm div}=[(\lambda_{\rm st}^{1/2}+\lambda_{\rm st'}^{1/2})^2-(\mst^2-\mst'^2)^2]/(4\msp^2)$ with
$\lambda_{\rm st}=\lambda(s,\mst^2,\msp^2)$ and $\lambda_{\rm st'}=\lambda(s,\mst'^2,\msp^2)$.

\section{Contact operators}
\label{app:fivepoint}

The magnetic operators of lowest order in derivatives contain one magnetic field $\vB$ and no further derivative.
Owing to HQSS, their spin structures act only on the light-quark spins, i.e., on the first index of $\Hb$, the
Cartesian index of $S^i$ and the spinor index of $N$. The photon couples as an isoscalar plus the third component
of an isovector. Since the model space contains three types of channels, $\Db^{(*)}\Sc^{(*)}$, $\Db^{(*)}\Lc$ and
the hidden-charm channels, there are six pairs of channel types. In terms of the light-spin bilinears
$\mathcal O^0_{\Hb}=\tr{\Hb^\dagger\Hb}$, $\mathcal O^j_{\Hb}=\tr{\Hb^\dagger\sigma^j\Hb}$,
$\mathcal O^0_S=S^{i\dagger}S^i$, $\mathcal O^j_S=i\epsilon^{ijk}S^{i\dagger}S^k$ and
$\mathcal O^{ij}_S=S^{i\dagger}S^j+S^{j\dagger}S^i-\tfrac23\delta^{ij}S^{k\dagger}S^k$, and of six isospin
structures labeled by $t$ for the $\Hb^\dagger\Hb S^\dagger S$ operators, the operators read
\begin{widetext}
\begin{equation}
\begin{aligned}
\mathcal L^{(1)}_\gamma&=eB^j\sum_{t\neq5}\Big\{d^{(a)}_t[\mathcal O^j_{\Hb}\otimes\mathcal O^0_S]_t+d^{(b)}_t[\mathcal O^0_{\Hb}\otimes\mathcal O^j_S]_t+d^{(c)}_t[\mathcal O^i_{\Hb}\otimes\mathcal O^{ij}_S]_t\Big\}
+e\,d^{(e)}_5\,\epsilon^{ijk}B^i[\mathcal O^j_{\Hb}\otimes\mathcal O^k_S]_5,\\
\mathcal L^{(2)}_\gamma&=eB^j\big\{d^{(d)}_0\tr{\Hb^\dagger\sigma^j\Hb}+d^{(d)}_1\tr{\Hb^\dagger\sigma^j\tau^3\Hb}\big\}\Lc^\dagger\Lc,\\
\mathcal L^{(3)}_\gamma&=eB^i\sum_{t=1}^3f_t\mathcal R^i_t+e\,i\epsilon^{ijk}B^i\sum_{t=1}^3f'_t\mathcal R^{jk}_t+{\rm h.c.},\\
\mathcal L^{(4)}_\gamma&=eB^j\sum_{t=1}^3\big\{h_tN^\dagger\sigma^i\sigma^j\Hb\Psi^\dagger S^i+h'_tN^\dagger\sigma^j\sigma^i\Hb\Psi^\dagger S^i\big\}_{\mathcal N_t}+{\rm h.c.},\\
\mathcal L^{(5)}_\gamma&=eB^jN^\dagger\sigma^j(k_0+k_1\tau^3)\Hb\Psi^\dagger\Lc+{\rm h.c.},\\
\mathcal L^{(6)}_\gamma&=eB^jN^\dagger\sigma^j(n_0+n_1\tau^3)N\,\tr{\Psi^\dagger\Psi},
\end{aligned}
\label{eq:L5}
\end{equation}
\end{widetext}
with
\begin{equation}
\begin{aligned}
\mathcal R^i_{1,2,3}&=\{S^{i,a\dagger}\tr{\Hb^\dagger\tau^a\Hb},\,S^{i,3\dagger}\tr{\Hb^\dagger\Hb},\\
&\qquad i\epsilon^{3ab}S^{i,a\dagger}\tr{\Hb^\dagger\tau^b\Hb}\}\Lc,\\
\mathcal N_{1,2,3}&=\{(N^\dagger\tau^a\Hb)S^a,\,(N^\dagger\Hb)S^3,\,i\epsilon^{3ab}(N^\dagger\tau^a\Hb)S^b\},
\end{aligned}
\label{eq:RN}
\end{equation}
where $\mathcal R^{jk}_t$ is defined as $\mathcal R^i_t$ with $\sigma^j$ inserted in the $\Hb$ trace, and all
couplings are real with dimension $E^{-4}$. The subscript $\mathcal N_t$ denotes the isospin contraction. The six
isospin structures are $\tr{\Hb^\dagger\Hb}S^\dagger\!\cdot S$, $V^a\mathcal I^a_S$, $V^3S^\dagger\!\cdot S$,
$\tr{\Hb^\dagger\Hb}\mathcal I^3_S$, $\epsilon^{3ab}V^a\mathcal I^b_S$ and $V^aQ^{a3}_S$ for $t=1,\ldots,6$, with
$V^a=\tr{\Hb^\dagger\tau^a\Hb}$, the isospin operator $\mathcal I^a_S$ of the $\Sc^{(*)}$ and
$Q^{ab}_S=\tfrac43\delta^{ab}-\{\mathcal I^a_S,\mathcal I^b_S\}$. Time-reversal invariance combines the spin
structures $a$, $b$ and $c$ with the five isospin structures that are even under time reversal, and the fourth spin
structure with $t=5$. This yields $16$ operators for the $\Db^{(*)}\Sc^{(*)}$ channels. Including the operators for
the other pairs of channel types, the total number is $16+2+6+6+2+2=34$. Two-body operators that act on the
heavy-quark spins are suppressed by $\Lambda_{\rm QCD}/m_c$ and are not included.

All operators have the form $e\,c\,B^j\mathcal X^j$, where $c$ is a coupling and $\mathcal X^j$ is a field
bilinear that transforms as a vector in the light-quark spins. For an incoming photon,
$A^\mu=\epsilon^\mu e^{-iq\cdot x}$ with $\epsilon^0=0$, one has $\vB=i\vq\times\bm\epsilon$, and
$\mathcal L_{\rm int}=-eJ^\mu A_\mu=e\bm J\cdot\bm\epsilon$ gives the current $\bm J=ic\,\bm{\mathcal X}\times\vq$.
This is the vertex \eqref{eq:GM} in the rest frame of the pair. Between the channel states of
Sec.~\ref{subsec:model}, the bilinears of $\mathcal L^{(1)}_\gamma$ and $\mathcal L^{(2)}_\gamma$ reduce to spin
operators of the light quark in the meson and of the light diquark,
\begin{equation}
\begin{aligned}
\bm{\mathcal X}^{(a)}&=\bm{\mathcal X}^{(d)}=2\vs_q,\qquad\bm{\mathcal X}^{(b)}=2\bm S_{qq},\\
\mathcal X^{(c)j}&=-2\,\sigma^i_q\,Q^{ij}_{qq},\\
Q^{ij}_{qq}&=S^i_{qq}S^j_{qq}+S^j_{qq}S^i_{qq}-\tfrac43\delta^{ij},
\end{aligned}
\label{eq:Xquark}
\end{equation}
up to isospin factors, which are absorbed into the combinations \eqref{eq:effLEC} below. The factor $2$ arises
from the trace over the spin of the $\bar c$. In the Condon--Shortley coupling scheme of Sec.~\ref{subsec:model},
Eq.~\eqref{eq:Xquark} reproduces the spin factors $\Xis^{(a)}$ to $\Xis^{(d)}$ of Table~\ref{tab:gamma}, including
their signs. In the $\Db^{(*)}\Sc^{(*)}$ channels, the magnetic vertex in the Breit frame is therefore
$\bm\Gt_M=4\sqrt{m_{l'}m_l}\,i\,[\bar d^{(a)}\vs_q+\bar d^{(b)}\bm S_{qq}-\bar d^{(c)}\vs_q\cdot\mathsf Q_{qq}]_{l'l}\times\vq$
with $(\vs_q\cdot\mathsf Q_{qq})^j=\sigma^i_qQ^{ij}_{qq}$. The operators of $\mathcal L^{(3)}_\gamma$ to
$\mathcal L^{(6)}_\gamma$ change the spin of the diquark from one to zero or connect the open-charm and
hidden-charm channels. Their spin factors were computed with the same fields and are listed in
Table~\ref{tab:gamma}. In the $I=\tfrac12$, $I_3=+\tfrac12$ state, the couplings enter through the combinations
\begin{equation}
\begin{aligned}
\bar d^{(s)}&=d^{(s)}_1-2d^{(s)}_2-\tfrac13d^{(s)}_3+\tfrac23d^{(s)}_4+\tfrac{20}9d^{(s)}_6,\\
\bar d^{(d)}&=d^{(d)}_0+d^{(d)}_1,\\
\bar f^{(\prime)}&=-\sqrt3f^{(\prime)}_1-\tfrac{1}{\sqrt3}f^{(\prime)}_2+\tfrac2{\sqrt3}f^{(\prime)}_3,\\
\bar n&=n_0+n_1,
\end{aligned}
\label{eq:effLEC}
\end{equation}
where $(s=a,b,c)$ and analogously $\bar h^{(\prime)}$ and $\bar k=k_0+k_1$. Since the structure $t=5$ has no diagonal matrix element
in an isospin eigenstate, $d^{(e)}_5$ does not contribute. The combinations $\bar d^{(a,b,c,d)}$, $\bar f$,
$\bar f'$, $\bar h$, $\bar h'$, $\bar k$ and $\bar n$ are the ten couplings $\bar c_s$ of
Sec.~\ref{subsubsec:contact}. The spin factors are listed in Table~\ref{tab:gamma}. For $J=\tfrac52$,
$\Xis^{(a)}=\Xis^{(b)}=\tfrac45$ and $\Xis^{(c)}=-\tfrac8{15}$.
\begin{table*}[t]
\caption{Spin factors $\Xis^{(s)}_{l'l}(J)$ of the lowest-derivative magnetic operators, Eq.~\eqref{eq:Jtr}. Bases:
$(\Db\Sc,\Dbs\Sc,\Dbs\Scs)$ for $J=\tfrac12$ and $(\Db\Scs,\Dbs\Sc,\Dbs\Scs)$ for $J=\tfrac32$. For $\Xis^{(d)}$ the
basis is $(\Db\Lc,\Dbs\Lc)$ for $J=\tfrac12$ and $(\Dbs\Lc)$ for $J=\tfrac32$. The rows (columns) of $\Xis^{(f)}$
and $\Xis^{(f')}$ are the $\Sc^{(*)}$ ($\Lc$) channels. The rows of $\Xis^{(h)}$, $\Xis^{(h')}$ and $\Xis^{(k)}$ are
the hidden-charm channels, $(\etac N,\Jpsi N)$ for $J=\tfrac12$ and $(\Jpsi N)$ for $J=\tfrac32$, and their columns
the $\Sc^{(*)}$ or $\Lc$ channels. $\Xis^{(n)}$ is diagonal in the hidden-charm channels. The matrices of the
transition operators are symmetric in the channel indices.}
\label{tab:gamma}
\begin{ruledtabular}
\begin{tabular}{c}
\begin{minipage}{0.97\textwidth}
\begin{equation*}
\begin{aligned}
J=\tfrac12:\quad
\Xis^{(a)}&=\begin{pmatrix}0&\tfrac4{\sqrt3}&0\\ \tfrac4{\sqrt3}&\tfrac83&0\\ 0&0&-\tfrac43\end{pmatrix},\qquad
\Xis^{(b)}=\begin{pmatrix}\tfrac83&0&0\\ 0&-\tfrac89&-\tfrac{8\sqrt2}9\\ 0&-\tfrac{8\sqrt2}9&\tfrac{20}9\end{pmatrix},\qquad
\Xis^{(c)}=\begin{pmatrix}0&0&\tfrac{20\sqrt6}9\\ 0&0&-\tfrac{20\sqrt2}9\\ \tfrac{20\sqrt6}9&-\tfrac{20\sqrt2}9&\tfrac{40}9\end{pmatrix},\\[4pt]
J=\tfrac32:\quad
\Xis^{(a)}&=\begin{pmatrix}0&0&\tfrac4{\sqrt{15}}\\ 0&\tfrac43&0\\ \tfrac4{\sqrt{15}}&0&\tfrac8{15}\end{pmatrix},\qquad
\Xis^{(b)}=\begin{pmatrix}\tfrac43&0&0\\ 0&\tfrac89&-\tfrac8{9\sqrt5}\\ 0&-\tfrac8{9\sqrt5}&\tfrac{44}{45}\end{pmatrix},\qquad
\Xis^{(c)}=\begin{pmatrix}0&-\tfrac4{3\sqrt3}&-\tfrac{16}{3\sqrt{15}}\\ -\tfrac4{3\sqrt3}&0&\tfrac{28}{9\sqrt5}\\ -\tfrac{16}{3\sqrt{15}}&\tfrac{28}{9\sqrt5}&\tfrac{64}{45}\end{pmatrix},
\end{aligned}
\end{equation*}
\begin{equation*}
\begin{aligned}
\Xis^{(d)}(\tfrac12)&=\begin{pmatrix}0&\tfrac4{\sqrt3}\\ \tfrac4{\sqrt3}&\tfrac83\end{pmatrix},\qquad
\Xis^{(d)}(\tfrac32)=\tfrac43,\qquad
\Xis^{(f)}(\tfrac12)=\begin{pmatrix}\tfrac4{\sqrt3}&0\\ 0&-\tfrac4{3\sqrt3}\\ 0&\tfrac{8\sqrt2}{3\sqrt3}\end{pmatrix},\qquad
\Xis^{(f')}(\tfrac12)=\begin{pmatrix}0&\tfrac83\\ -\tfrac83&0\\ \tfrac{4\sqrt2}3&\tfrac{4\sqrt6}3\end{pmatrix},\\[4pt]
\Xis^{(f)}(\tfrac32)&=\begin{pmatrix}0\\ \tfrac4{3\sqrt3}\\ \tfrac8{3\sqrt{15}}\end{pmatrix},\qquad
\Xis^{(f')}(\tfrac32)=\begin{pmatrix}-\tfrac43\\ 0\\ \tfrac4{\sqrt{15}}\end{pmatrix},\qquad
\Xis^{(k)}(\tfrac12)=\begin{pmatrix}2&2\sqrt3\\ -\tfrac2{\sqrt3}&\tfrac23\end{pmatrix},\qquad
\Xis^{(n)}(\tfrac12)=\begin{pmatrix}4&0\\ 0&-\tfrac43\end{pmatrix},
\end{aligned}
\end{equation*}
\begin{equation*}
\begin{aligned}
\Xis^{(h)}(\tfrac12)&=\begin{pmatrix}-\tfrac2{\sqrt3}&\tfrac23&-\tfrac{4\sqrt2}3\\ \tfrac{10}3&\tfrac{14}{3\sqrt3}&-\tfrac{4\sqrt2}{3\sqrt3}\end{pmatrix},\qquad
\Xis^{(h')}(\tfrac12)=\begin{pmatrix}2\sqrt3&-2&4\sqrt2\\ \tfrac23&-\tfrac{10}{3\sqrt3}&-\tfrac{4\sqrt2}{3\sqrt3}\end{pmatrix},\\[4pt]
\Xis^{(h)}(\tfrac32)&=\begin{pmatrix}\tfrac43&\tfrac4{3\sqrt3}&-\tfrac4{3\sqrt{15}}\end{pmatrix},\qquad
\Xis^{(h')}(\tfrac32)=\begin{pmatrix}-\tfrac43&\tfrac4{3\sqrt3}&\tfrac{4\sqrt{15}}9\end{pmatrix},\qquad
\Xis^{(k)}(\tfrac32)=\Xis^{(n)}(\tfrac32)=\tfrac43,
\end{aligned}
\end{equation*}
\end{minipage}
\end{tabular}
\end{ruledtabular}
\end{table*}

The size of the couplings can be estimated within the vector-exchange model. Since the photon couples to the
exchanged $\rho$ only through the isospin structure $t=5$ and the leading vector-meson couplings are spin
independent, the leading remaining mechanism is the anomalous $\pi V\gamma$ exchange current. In this mechanism,
one hadron emits a pion, the photon converts it into a $\rho$ or $\omega$ meson, and the second hadron absorbs the
vector meson. We take the $\pi V\gamma$ couplings from the experimental radiative widths of the $\omega$ and the
$\rho$, the pion couplings from $\Gamma(D^{*+})$ and $\Gamma(\Sc^{++}\to\Lc^+\pi^+)$ and the vector coupling
from the normalization of the kernel. The combinations \eqref{eq:effLEC} are then $9$ to $12\GeV^{-4}$ in
modulus for $\bar d^{(a)}$, $\bar d^{(d)}$ and $\bar f$ and below $1\GeV^{-4}$ for $\bar d^{(b)}$. Couplings of this size
change the magnetic moments by at most $0.07\,\muN$. Since the exchange current has the range of the pion and the
relative signs of the couplings were not determined, this provides only an order-of-magnitude estimate.

The electric operators of Eq.~\eqref{eq:LE} contain a two-hadron operator $\mathcal X$ that is a Hermitian scalar
in the light-quark spins and even under time reversal. For the $\Db^{(*)}\Sc^{(*)}$ channels, the scalar spin
structures are $\mathcal O^0_{\Hb}\mathcal O^0_S$ and $\mathcal O^j_{\Hb}\mathcal O^j_S$, which are equivalent to
one coupling for each light-quark spin, as in the case of the kernel. Combined with the five even isospin
structures, they give $10$ operators. With one scalar spin structure for each of the other pairs of channel types,
the total number is $10+2+3+3+2+2=22$. In an $I=\tfrac12$ state, each scalar spin structure of each pair of channel
types contributes through one combination, which yields the seven couplings of the matrix $\Ecal$. Between the
channel states, the two scalar structures reduce to $2$ and $2\vs_q\cdot\bm S_{qq}$. With
$\bm\nabla\cdot\vE=v_\nu\partial_\mu F^{\mu\nu}$ and $\partial_\mu\to-iq_\mu$ for the incoming photon, one has
$\partial_\mu F^{\mu\nu}\to-q^2\epsilon^\nu+q^\nu(q\cdot\epsilon)$, and $\mathcal L_{\rm int}=-eJ^\mu A_\mu$ gives
$J^\mu=d\,[q^2v^\mu-q^\mu(q\cdot v)]\,\mathcal X$, which is Eq.~\eqref{eq:GE}. In the Breit frame at $s=s'$, the
charge density is $-d\,Q^2\mathcal X$, and $\MSR=-6\,\dd F/\dd Q^2$ leads to Eq.~\eqref{eq:drE}.

\section{One-body magnetic matrix elements}
\label{app:moments}

Table~\ref{tab:inventory} lists the reduced matrix elements \eqref{eq:WE} of the one-body magnetic operator
between the channel states, separated into the meson-line and baryon-line contributions. They are computed with
the spin wave functions of the multiplet fields of Sec.~\ref{subsec:model} and the moments of
Table~\ref{tab:moments}. For two spins $j_1$ and $j_2$ coupled to $J$,
$\langle\bm j_1\rangle=[J(J+1)+j_1(j_1+1)-j_2(j_2+1)]/[2J(J+1)]\,\vJ$, which yields the additive moments on the
diagonal. In the $\Sc^{(*)}$ channels, the meson moments are averaged with the isospin weights $\tfrac23$ and
$\tfrac13$, while in the $\Lc$ channels the meson is a $\Db^{(*)0}$. In the hidden-charm channels, the baryon line
contributes $\mu_p$ for $\etac N$ with $J=\tfrac12$ and for $\Jpsi N$ with $J=\tfrac32$, and $-\mu_p/3$ for
$\Jpsi N$ with $J=\tfrac12$, whereas the meson line contributes $0.482\,\muN$ to the transition between $\etac N$
and $\Jpsi N$.

The transition moments of the mesons can be compared with experimental data. With
$\Gamma(V\to P\gamma)=(\alpha/3)(k^3/M_p^2)|\mu_{VP}/\muN|^2$, where $k$ is the photon momentum, the experimental
widths~\cite{ParticleDataGroup:2026} yield $|\mu(D^{*+}\to D^+\gamma)|=0.44\pm0.06\,\muN$ and
$|\mu(\Jpsi\to\etac\gamma)|=0.67\pm0.03\,\muN$, which are $14$ and $20\%$ below the constituent-quark values
$0.51$ and $0.83\,\muN$, respectively.

\begin{table*}[t]
\caption{Isospin-averaged one-body magnetic matrix elements $\mathcal M^{(1)}_{l'l}(J)$ in $\muN$ between the
open-charm channel states of the $P_c^+$. Each entry gives the meson-line part $\mathcal M^{\rm M}$ / the
baryon-line part $\mathcal M^{\rm B}$.}
\label{tab:inventory}
\begin{ruledtabular}
\begin{tabular}{l|ccccc}
$J=\tfrac12$ & $\Db\Sc$ & $\Dbs\Sc$ & $\Dbs\Scs$ & $\Db\Lc$ & $\Dbs\Lc$\\
\hline
$\Db\Sc$   & $0$ / $+1.723$ & $+0.241$ / $0$ & $0$ / $0$ & $0$ / $-0.931$ & $0$ / $0$\\
$\Dbs\Sc$  & $+0.241$ / $0$ & $-0.278$ / $-0.574$ & $0$ / $-0.615$ & $0$ / $0$ & $0$ / $+0.310$\\
$\Dbs\Scs$ & $0$ / $0$ & $0$ / $-0.615$ & $+0.139$ / $+1.783$ & $0$ / $0$ & $0$ / $-0.878$\\
$\Db\Lc$   & $0$ / $-0.931$ & $0$ / $0$ & $0$ / $0$ & $0$ / $+0.417$ & $+1.316$ / $0$\\
$\Dbs\Lc$  & $0$ / $0$ & $0$ / $+0.310$ & $0$ / $-0.878$ & $+1.316$ / $0$ & $+0.963$ / $-0.139$\\
\hline
$J=\tfrac32$ & $\Db\Scs$ & $\Dbs\Sc$ & $\Dbs\Scs$ & $\Dbs\Lc$ & \\
\hline
$\Db\Scs$  & $0$ / $+3.209$ & $0$ / $0$ & $+0.323$ / $0$ & $0$ / $0$ & \\
$\Dbs\Sc$  & $0$ / $0$ & $-0.417$ / $+1.723$ & $0$ / $-0.584$ & $0$ / $-0.931$ & \\
$\Dbs\Scs$ & $+0.323$ / $0$ & $0$ / $-0.584$ & $-0.167$ / $+2.354$ & $0$ / $-0.833$ & \\
$\Dbs\Lc$  & $0$ / $0$ & $0$ / $-0.931$ & $0$ / $-0.833$ & $+1.445$ / $+0.417$ & \\
\hline
$J=\tfrac52$ & $\Dbs\Scs$ & & & & \\
\hline
$\Dbs\Scs$ & $-0.417$ / $+3.209$ & & & & \\
\end{tabular}
\end{ruledtabular}
\end{table*}

\end{document}